\documentclass[preprintnumbers,nofootinbib,amsmath,amssymb,twocolumn]{revtex4}

\usepackage{amsmath}
\usepackage{amsfonts}
\usepackage{amssymb}
\usepackage{graphicx}    
\usepackage{amsfonts}
\usepackage{amsmath}
\usepackage{braket}
\usepackage{rotating}
\usepackage{verbatim}
\usepackage{multirow}
\usepackage{hyperref}
\newcounter{acounter}
\newtheorem{assumption}{Assumption}{\bfseries}{\itshape}

\newcommand{\planck}{{\sc Planck}}

\newcommand{\WMAP}{{\sc WMAP}}

\newcommand{\vect}[1]{\ensuremath{\boldsymbol{#1}}}

\newcommand{\fnl}{f_{\rm NL}^\bigtriangleup}

\newcommand{\vk}{\vect{k}}

\newcommand{\flow}[1]{{^{#1}\!\lambda}}
\newcommand{\vev}[1]{\langle #1\rangle}

\begin{document}

\preprint{Imperial/TP/2013/CC/3} \vskip 0.2in

\title{\planck\ and \WMAP\ constraints on generalised Hubble flow inflationary
  trajectories}


\author{Carlo~R.~Contaldi} \email{c.contaldi@imperial.ac.uk}
\affiliation{Theoretical Physics, Blackett Laboratory, Imperial College, London, SW7 2BZ, UK}
\affiliation{Canadian Institute of Theoretical Physics, 60 St. George
  Street, Toronto, M5S 3H8, On, Canada}
\author{Jonathan~S.~Horner}
\affiliation{Theoretical Physics, Blackett Laboratory, Imperial College, London, SW7 2BZ, UK}

\date{\today}

\begin{abstract}
  We use the Hamilton--Jacobi formalism to constrain the space of
  possible single field, inflationary Hubble flow trajectories when
  compared to the \WMAP\ and \planck\ satellites Cosmic Microwave Background (CMB)
  results. This method yields posteriors on the space of Hubble Slow
  Roll (HSR) parameters that uniquely determine the history of the
  Hubble parameter during the inflating epoch. The trajectories are
  used to numerically determine the observable primordial power
  spectrum and bispectra that can then be compared to
  observations. Our analysis is used to infer the most likely shape of
  the inflaton potential $V(\phi)$ and also yields a prediction for, $f_{\rm
    NL}$, the dimensionless amplitude of the non-Gaussian bispectrum.
\end{abstract}


\maketitle

\section{Introduction}\label{sec:intro}

Recent \planck\ results \cite{2013arXiv1303.5062P} have confirmed, with the
highest precision to date, the existence of a spectrum of primordial
curvature perturbations on super-horizon scales with a power law with
a spectral index close to but not equal to unity. This picture has now
been verified over roughly three decades of scales probed by primary
Cosmic Microwave Background (CMB) anisotropies that can be related to
the primordial curvature perturbations on super-horizon scales via a
well defined set of photon perturbation transfer functions. The
quoted value for the scalar spectral index of $n_s = 0.9603\pm 0.0073$
seems to be in good agreement with many models of cosmological
inflation \cite{2013arXiv1303.5082P}. The fact that $n_s$ is not compatible
with unity is also interpreted by many to support the actual {\sl
  existence} of an inflationary epoch in the very early universe.

The interpretation of the result, in the context of inflationary model
selection, is complicated by the large number of inflationary models
that are compatible with the CMB observations (see for example section~2
of \cite{2013arXiv1303.5082P} for a review of the landscape of models). The
models range from the simplest chaotic model with a single scalar
field  to massively, multi-field models inspired by
dimensional compactification in string theory. A
typical discriminatory approach is to analyse the consistency of a
particular model in the space of parameters such as $n_s$ and
tensor-to-scalar ratio $r$ constrained directly by the data. However
it becomes readily apparent that this combination does not refine the
space of possible models to an extent at which conclusions about the
fundamental nature of the inflaton can be made. Including higher-order
parameters (in either slow-roll approximation or perturbation
expansion) such as running of the spectral index with wavenumber $k$
$dn_s/d\ln k$ or non-Gaussianity amplitude $f_{\rm NL}$ greatly
enhances the ability to reject or falsify models. However the data has
not reached the sensitivity to detect the expected higher-order
signals.

An alternative method adopted here is known as the Hubble flow
equation method. This method \cite{Kinney:1997ne,Easther:2002rw,Liddle:2003py,2005PhRvD..72h3520C} assumes inflation was
driven by a single scalar field and employs the Hamilton-Jacobi
framework \cite{PhysRevD.42.3936} to define a hierarchy of
differential equations that can be used to generate inflationary trajectories consistent
with {\sl any} inflationary potential up to a certain order in derivatives of
the Hubble rate $H$ with respect to the inflaton field value $\phi$.
Within this framework one can dispense with proposing a {\sl single}
model consisting of a parametrised potential and constrain directly
the space of allowed inflationary trajectories described by the
evolution of the Hubble parameter $H\equiv \dot a/a$.

This approach allows one to compare all possible inflationary
trajectories with a given complexity with no loss of accuracy. This is
because, for a given truncation of the hierarchy of differential
equations, the value of the Hubble rate during the inflationary epoch
can be evaluated to arbitrary precision. Once the history of $H$ has
been obtained it is then possible to calculate all observable
quantities to the desired precision irrespective of whether the
trajectory satisfies slow-roll conditions.

One can then use the hierarchy of Hubble flow parameters as the base
parameters being constrained. This has two advantages. Firstly, the
space of Hubble flow parameters explores the space of all inflationary
potentials allowed at a certain order consistently. Secondly, Bayesian
model comparison is simplified for a given Hubble flow order since
there is a {\sl single} model proposition which reduces the selection
to a comparison of likelihood values for two different points in the
Hubble parameter space with no need to calculate Bayesian evidence.

In the work reported here \planck\ total intensity results, together with \WMAP\
polarisation results \cite{2013ApJS..208...20B} are used to directly constrain the
space of Hubble flow parameters with priors given by the set of
Assumptions~\ref{a:exist}-\ref{a:phi}. The constrained space of Hubble flow parameters can
then be related to ``conventional'' parameters including $n_s$, $r$,
$dn_s/d\ln k$, and $f_{\rm NL}$ without the need to redefine the model
from a Bayesian perspective. The definition of {\sl detections} of the
conventional parameters $n_s-1$, $r$, $dn_s/d\ln k$, and $f_{\rm NL}$
has no meaning within this analysis and the constraints can be viewed
as ranges {\sl allowed} by the observations i.e. {\sl predictions}
given the underlying set of assumptions.

This {\sl paper} is organised as follows. In section~\ref{sec:rev} we
review the Hubble flow formalism and describe how to obtain
observables to compare with data in section~\ref{sec:calc}. In
section~\ref{sec:res} we show the results obtained by constraining
Hubble flow trajectories using the latest CMB data. We also describe
the derived constraints on primordial spectral parameters and on the
inflaton potential. In section~\ref{sec:disc} we discuss our results
and future extensions.


\section{Hubble flow equations}\label{sec:rev}

The Hamilton-Jacobi approach to analysing the dynamics of inflation
consists of changing the independent variable in the Friedmann equations
from cosmological time $t$ to the value of the inflaton scalar field
$\phi$. The only assumption required for this change of variable is
that $\phi$ is a monotonic function of $t$.  The Friedmann equation
and the inflaton's equation of motion then take on the following form
\begin{eqnarray}
  \dot{\phi} & = & -2M^{2}_{pl}H'(\phi)\,,\\
  \left[H'(\phi)\right]^{2} - \frac{3}{2M^{2}_{pl}}H(\phi)^{2} & = & -\frac{1}{2M^{4}_{pl}}V(\phi)\,,
  \label{Friedmann}
\end{eqnarray}
where dot denotes a derivative with respect to $t$, prime denotes a
derivative with respect to $\phi$, $H\equiv\dot a/a$ is the Hubble
rate for the FRW scale factor $a(t)$, $M_{pl}$ is the Planck mass and
$V(\phi)$ is the inflaton potential. One of the advantages of
performing this change of of variables is that one can merely pick a
function $H(\phi)$ and this will correspond to an exact solution of a
corresponding potential $V(\phi)$ in (\ref{Friedmann}).

The system can be further simplified by introducing an infinite
hierarchy of Hubble flow parameters \footnote{Sometimes called
  Hubble-Slow-Roll (HSR) parameters to contrast with the
  Potential-Slow-Roll (PSR) parameters.}
\begin{equation}\label{lambda}
  \flow{\ell} = \left(2M^{2}_{pl}\right)^{\ell} \frac{(H')^{\ell-1}}{H^{\ell}}\frac{d^{(\ell+1)}H}{d\phi^{(\ell+1)}}\,.
\end{equation}
The first of these parameters, $\flow{0}\equiv \epsilon$, is a proxy
for the acceleration of the scale factor and it is straightforward to
verify that the relation
\begin{equation}\label{epsilon}
  \epsilon = 2M^{2}_{pl}\left(\frac{H'(\phi)}{H(\phi)}\right)^{2} = \frac{-\dot{H}}{H^{2}} = \frac{\dot{\phi}^{2}}{2M_{pl}^{2}H^{2}} < 1\,,
\end{equation}
is a necessary and sufficient condition for the universe to be
undergoing inflation with $\ddot a/a>0$ \footnote{This is in contrast
  to the the PSR $\epsilon_V$ for which $\epsilon_V < 1$ is only an
  {\sl approximate} condition for inflation.}. The $\ell=1$ and $2$
flow parameters can also be identified with the usual slow roll
parameter $\eta = \flow{1} = -(\ddot{\phi}/H\dot{\phi})$ and $\xi =
\flow{3}$.

A further change of variable can be introduced by using the relation
between the rate of change in $e$-folds $N=\ln(a/a_i)$, where $a_i$ is
the value of the scale factor at the beginning of inflation, and
cosmological time $t$ with $dN/dt=H$. The entire system can then be
re-cast as an infinite hierarchy of differential ``Hubble flow''
equations with $N$ as the independent variable
\begin{eqnarray}\label{background}
  \frac{\mathrm{d}H}{\mathrm{d}N} & = & -\epsilon\,H\,,\\
  \frac{\mathrm{d}\epsilon}{\mathrm{d}N} & = & 2\,\epsilon\,(\epsilon - \eta)\,,\\
  \frac{\mathrm{d}\flow{\ell}}{\mathrm{d}N} & = & \left[\ell\,\epsilon - (\ell - 1)\,\eta\right]\,\flow{\ell} - \flow{\ell+1}\,,
\end{eqnarray}
with solutions $H(N)$ and $\flow{\ell}(N)$.

This is the most natural set of variables to use when constructing
single field inflationary trajectories and the solution of the
infinite system provides a {\sl complete} set of exact solutions for
the background evolution that are consistent with single field
inflation and monotonic time evolution of $\phi$. Truncating the
hierarchy at $\ell_{\rm max}$ provides an {\sl incomplete} set of
solutions that are nonetheless still {\sl exact}.

In practice a set of solutions for a given $\ell_{\rm max}$ is
obtained by integrating the system with a set of random initial
conditions for $H$ and $\flow{\ell}$ for $\ell=0,1,...,\ell_{\rm
  max}$. The system can be integrated forward or backwards to obtain
an exact solution describing the dynamics of the background within a
required window in $e$-foldings $N$.
\begin{figure*}
  \centering
  \begin{tabular}{cc}
    \includegraphics[width=3.5in]{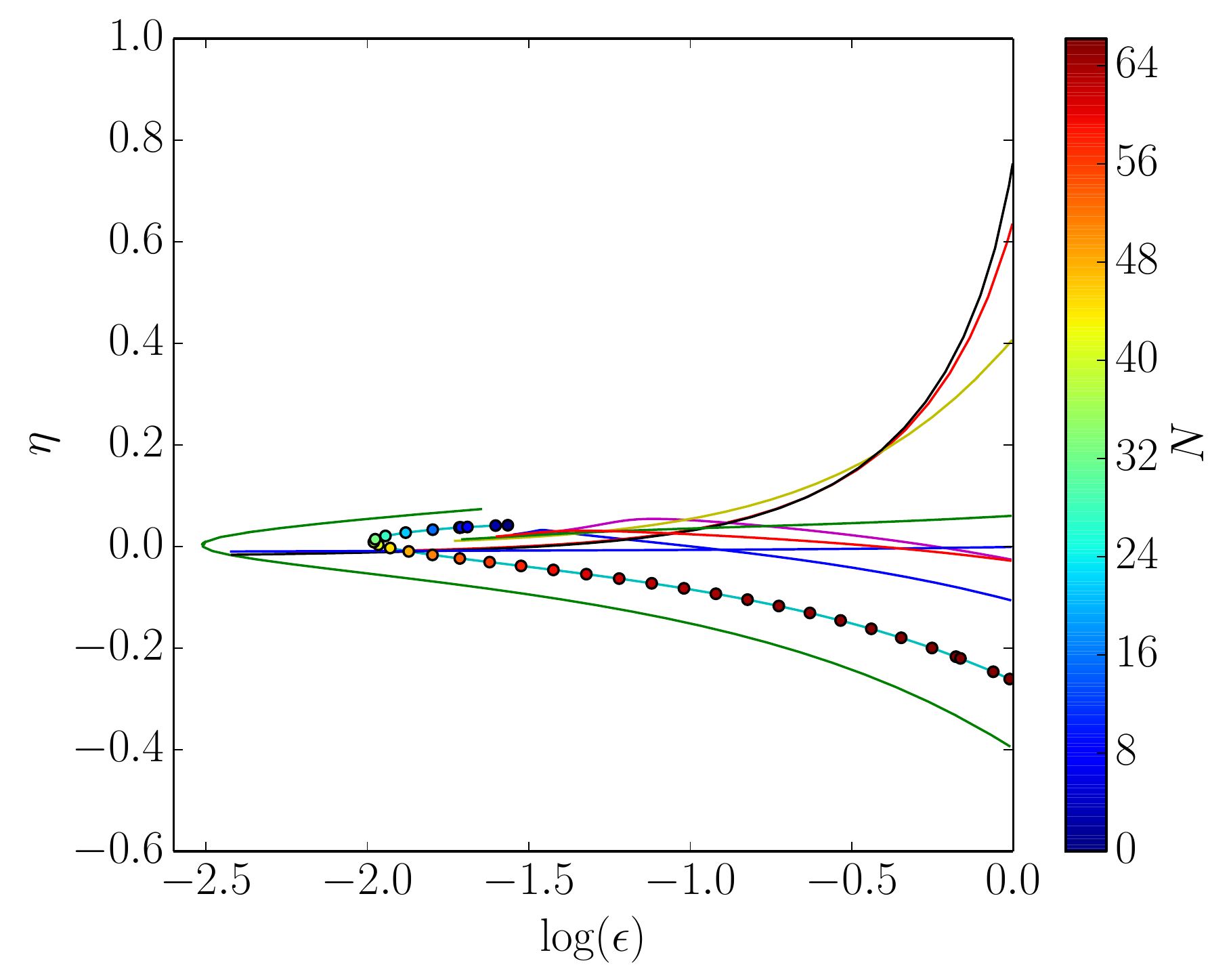}&
    \includegraphics[width=3.5in]{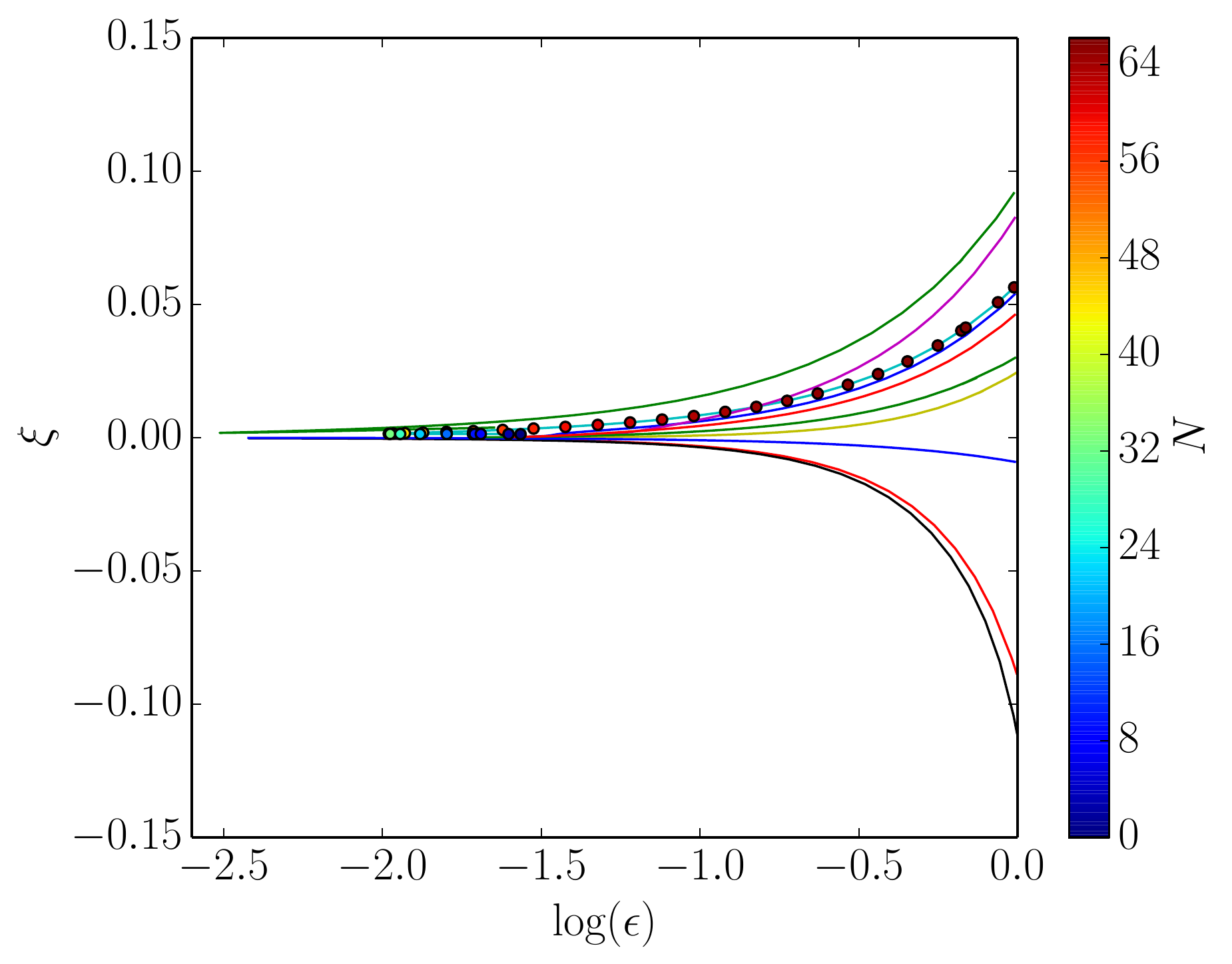}
    \end{tabular}
  \caption{Ten random trajectories drawn using the scheme described in
    (\ref{uniform}). The evolution of $\eta$ (left) and $\xi$ (right) are plotted against
    $\log(\epsilon)$ from the end of inflation ($\epsilon=1$) back to
    a time when the largest scale of interest $k_\star$ was a few order of
    magnitude smaller than the horizon scale. One of the trajectories
    also shows points colour coded by $e$-folding number $N$ as points colour coded with respect to $e$-fold
    $N$. $N=0$ corresponds a few $e$-folds before the $k_\star$ exits the horizon.  All
    trajectories are evolving away (as $N$ increases) from a
    ``slow-roll'' attractor with $\epsilon\ll 1$, $\eta\ll 1$, and
    $\xi\ll 1$. For most trajectories observable scales exit the
    horizon, when $N \sim {\cal O}(1) \rightarrow {\cal O}(10)$ and
    the flow parameters are well within the slow-roll
    limit. Trajectories with larger, negative final $\eta$ values are
    ones where the trajectory is furthest from the slow-roll regime at
  early times.}
  \label{fig:traj}%
\end{figure*}

\subsection{Hubble flow measure}\label{measure}
The Hubble flow method of generating random inflationary trajectories
has a well known measure problem due to the seemingly arbitrary
choice of proposal density and location for the initial conditions in
$H$ and $\flow{\ell}$.  The existence of attractors in the phase space
of the $\flow{\ell}$ complicates the interpretation of the imposed
measure and the nature of trajectories obtained.

A number of choices have been made in the literature
\cite{Kinney:1997ne,Easther:2002rw,Liddle:2003py,2005PhRvD..72h3520C}. These
include starting at arbitrary points and integrating forward or
backwards to select trajectories with enough $e$-folds. Different
choices have been made with regards to the encounters with fixed
points in the HSR phase space where $\epsilon$ asymptotes to a
constant and $^\lambda\ell \to 0$ for $\lambda > 1$. These can be
interpreted as eternally inflating solutions that can be allowed or
discarded if only trajectories where inflation ends are to be
allowed. In all cases the proposal densities for the HSR have been
uniform and the random draws have been made {\sl wherever} each trajectory's
integration was started. 

In this work the simplest possible assumptions compatible with the
data are made to define the choice of location for the initial
conditions
\begin{assumption}
  \label{a:exist}
  A phase of accelerated expansion (inflation) with $\ddot a > 0$
  occurred before the radiation dominated, decelerating phase of the
  standard big bang model.
\end{assumption}
\begin{assumption}
  \label{a:N}
  Inflation lasted a minimum number of $e$-folds such that all scales
  that are sub-horizon sized today were super-horizon by the end of
  inflation.
\end{assumption}
\begin{assumption}
  \label{a:end}
  Inflation ended when the universe stopped accelerating i.e. $\ddot
  a$ switched sign.
\end{assumption}
\begin{assumption}\label{a:phi}
  Inflation was driven by a single scalar field $\phi$.
\end{assumption}

In line with these assumptions the initial conditions are drawn at the
{\sl end} of inflation i.e. a fixed point where $\epsilon=1$. Only the
remaining flow parameters for $\ell=1,..,\ell_{\rm max}$ are then
drawn from uniform distributions with fixed ranges. A value for $N_0$
is drawn from a uniform distribution and the system (\ref{background})
is integrated {\sl backwards} a total number of $e$-folds $N_0$. The
$e$-folding $N_0$ is interpreted as the $e$-folding where the largest
mode $k_0$ in the observable window is sufficiently smaller than the
horizon to allow normalisation using the Bunch-Davies adiabatic limit
\cite{Bunch:1978yq}. This ensures that the system is integrated far back
enough for the calculation of all observables required for comparison
with data.  In all cases considered in this work $N_0$ is drawn with a
uniform distribution in the range $N_0=[60,70]$, this allows for the
uncertainty in the total number of $e$-folds that occurred after the
end of inflation due to the details of reheating. The uncertainty
impacts our ability to connect a given scale exiting the horizon at a
given time during inflation with a physically observable scale that
subsequently re-entered the horizon during the decelerating epoch (see
e.g. equation~(24) of \cite{2013arXiv1303.5082P}).

\subsection{Potential reconstruction}

Each trajectory generated in this manner corresponds to a realisation
of inflation with particular initial conditions and potential
$V(\phi)$. Given a trajectory one can reconstruct the potential
function probed during the evolution as the solution to the Hubble
flow system is equivalent to selecting a solution by specifying a
potential $V(\phi)$ and initial conditions for $\phi$ and
$\dot{\phi}$.

For example if $\flow{\ell} = 0$ for all $\ell > 0 $ then the only
remaining non-zero parameter is $\epsilon$. This implies $H(\phi)$ is
a linear function and hence $V(\phi)$ is quadratic. The solutions for
$\epsilon(N)$ and therefore $H(N)$ and $\phi(N)$ can then be obtained
easily. The potential is obtained by combining (\ref{Friedmann}) and
(\ref{epsilon}) to get
\begin{equation}\label{potential}
  V[\phi(N)] = 3M^{2}_{pl}H^{2}(N)\left[ 1 - \frac{\epsilon(N)}{3}\right]\,.
\end{equation}

\section{Calculation of observables}\label{sec:calc}

\begin{figure*}
  \centering
  \begin{tabular}{ll}
    \includegraphics[width=3.0in]{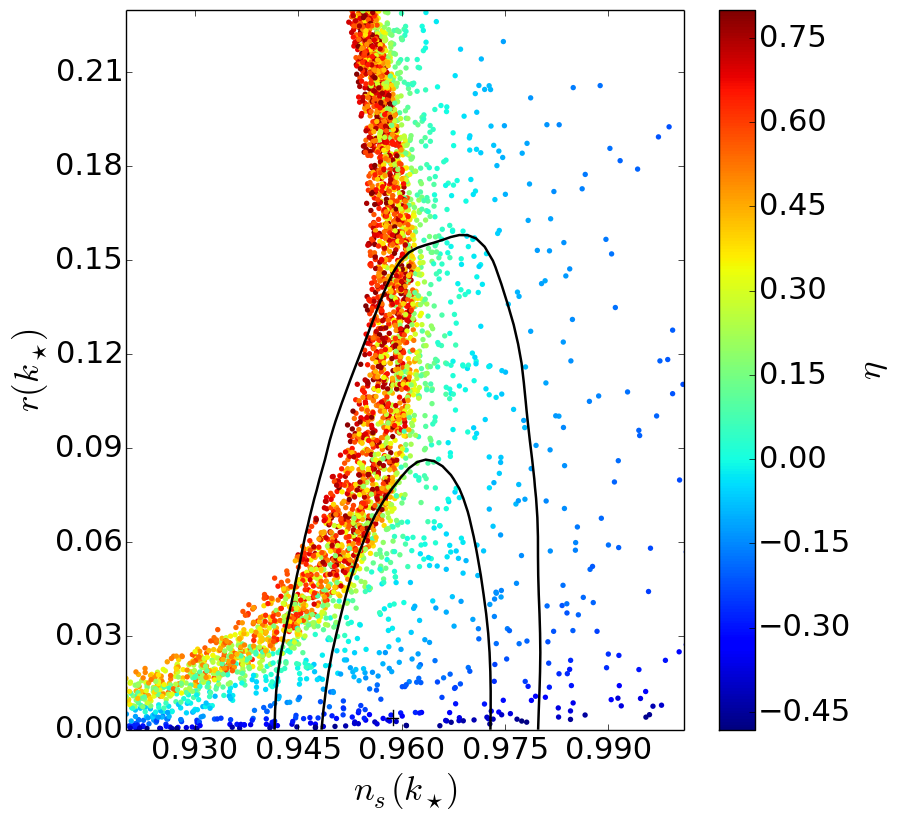}&
    \includegraphics[width=3.0in]{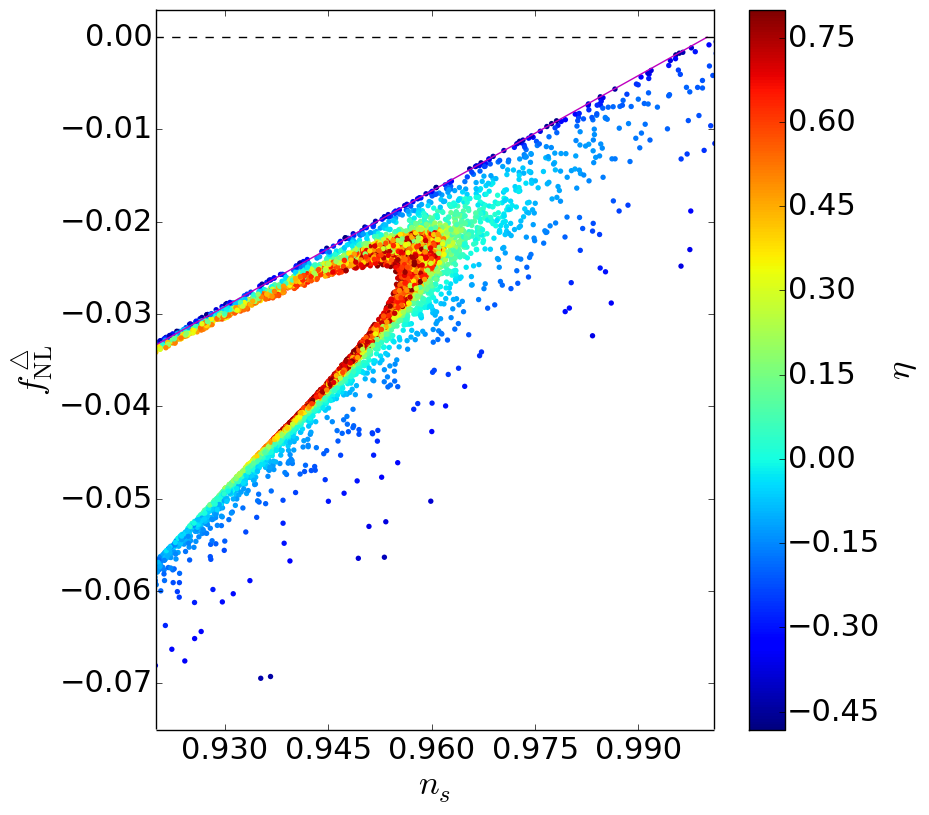}\\
    \includegraphics[width=3.0in]{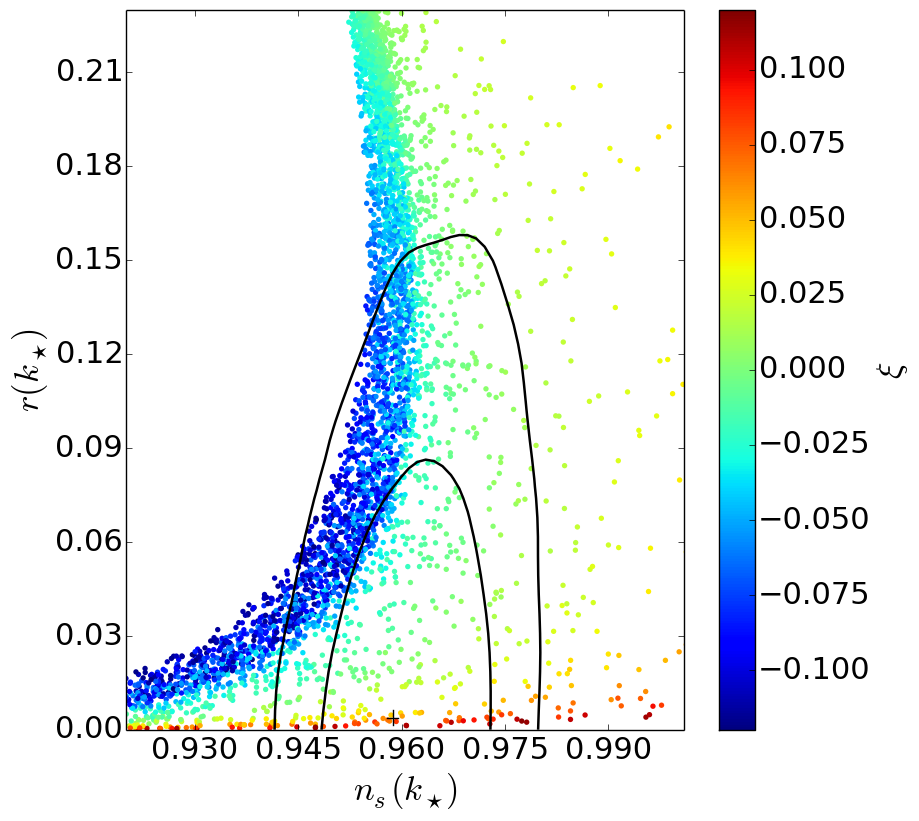}&
    \includegraphics[width=3.0in]{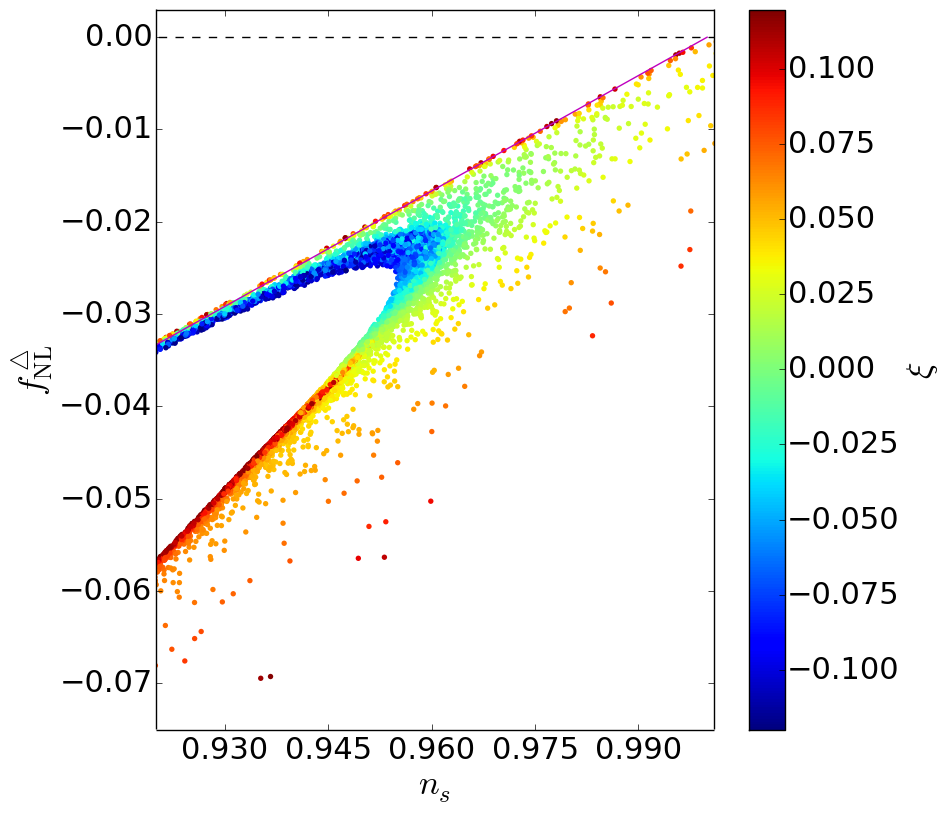}\\
    \includegraphics[width=3.0in]{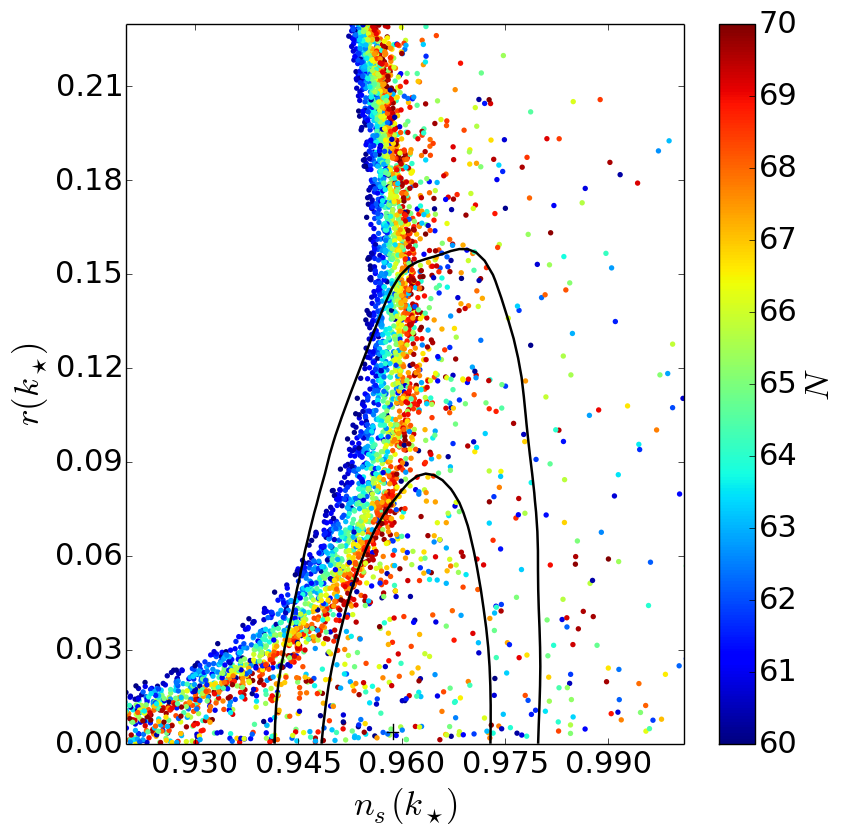}&
    \includegraphics[width=3.0in]{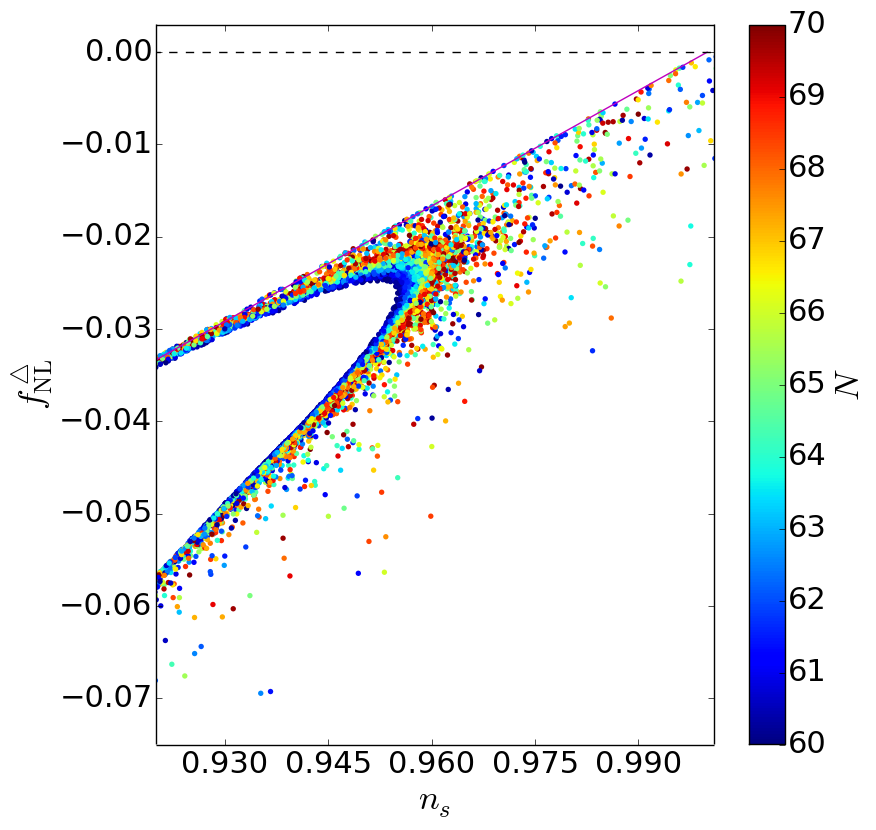}
  \end{tabular}	
  \caption{Hubble flow proposal densities projected into the space of
    derived parameters $n_s$, $r$, and $\fnl$. The derived parameters
    are obtained from numerical calculation of scalar and tensor power
    spectra and $\fnl$ and evaluated at the pivot scale $k_\star=0.05$
    Mpc$^{-1}$. The contours indicate the 68\% and 95\% confidence
    regions in the $n_s$-$r$ plane from the PLANCK$r$ reference fits
    \cite{2013arXiv1303.5076P}. Each point represents the derived quantities
    at the pivot scale obtained for each of $\sim 10000$ random Hubble
    flow trajectories generated using the uniform sampling described
    in section~\ref{measure}. The points are colour coded according to
    the random values $\eta$, $\xi$, and $N_0$ used to generate the
    trajectory. The inflationary attractor and the level of its
    correlation to the underlying flow parameters is clearly visible
    in both $n_s$-$r$ and $n_s$-$\fnl$ planes.  The $\fnl$
      attractor follows a consistency relation given by $f_{\rm
        NL}\sim \frac{5}{12}(n_s-1)$ 
      \cite{2013arXiv1303.2119H,2013arXiv1311.3224H} shown as the solid (magenta) line.}
  \label{fig:proposal}%
\end{figure*}

\begin{table*}
  \caption{Uniform MCMC priors for cosmological parameters and a short
    description of each parameter. \planck\ Nuisance parameters are not listed here but are included with the same prior settings as used in \cite{2013arXiv1303.5076P}. The second block are derived parameters that are not used to randomly sample trajectories.}  
  \label{table:params}  
  \centering                         
  \begin{tabular}{lcl}       
    \hline\hline                
    Parameter & Prior range & Definition \\    
    \hline                        
    \\[0.1cm]
    $\omega_b\equiv \Omega_b\,h^2$ & [0.005,0.1]& Baryon density today\\
    $\omega_c\equiv \Omega_c\,h^2$ & [0.001,0.99]& Cold dark matter density today\\
    $100\, \theta_{MC}$ & [0.5,10.0]& 100 $\times$ {\tt CosmoMC} sound horizon to angular diameter distance ratio approximation\\
    $\tau$ & [0.01,0.8]& Optical depth to reionisation\\
    $\ln(\tilde H_{\rm inf})$ & [2.5,3.5]& Log of rescaled
    Hubble rate at time of Horizon exit of scale $k_\star$\\
    $N_0$ &[60,70]& Number of $e$-folds for which trajectory is integrated back from end of inflation\\
    $\flow{0}\equiv \epsilon$& 1.0& Flow parameter value at end of inflation\\
    $\flow{1}\equiv \eta$& [-1.0,1.0]& Flow parameter value at end of inflation\\
    $\flow{2}\equiv \xi$& [-0.2,0.2]& Flow parameter value at end of inflation\\[0.1cm]
    \hline\\[0.1cm]
    $n_s(k_\star)$&... & Scalar spectral index measured from trajectory spectrum at scale $k_\star=0.05$ Mpc$^{-1}$\\
    $r(k_\star)$& ...& Tensor-to-scalar ratio measured from trajectory spectra at scale $k_\star=0.05$ Mpc$^{-1}$\\
    $n_t(k_\star)$& ...& Tensor spectral index measured from trajectory spectrum at scale $k_\star=0.05$ Mpc$^{-1}$\\
    $\fnl(k_\star)$ &...& Equilateral non-Gaussianity amplitude at scale $k_\star=0.05$ Mpc$^{-1}$\\
    $\epsilon(k_\star)$ &...& Flow parameter value shortly after mode $k_\star$ exits the horizon\\
    $\eta(k_\star)$ &...& Flow parameter value shortly after mode $k_\star$ exits the horizon\\
    $\xi(k_\star)$ &...& Flow parameter value shortly after mode $k_\star$ exits the horizon
    \\[0.1cm]
    \hline                                   
  \end{tabular}
\end{table*}

\subsection{Power spectrum}
The evolution of background, homogeneous quantities during inflation
is fully determined by the Hubble flow trajectory. Background also
determines the evolution of the inflaton perturbations that end up as
super-horizon primordial curvature perturbations that seed structure
formation after inflation. The power spectrum of primordial curvature
perturbations can be calculated numerically for any given Fourier
wavenumber $k\equiv |\vk|$. This is done by integrating the
Mukhanov-Sasaki \cite{mukhanov1982vacuum,Sasaki:1986hm} equation for the Fourier
expanded comoving curvature perturbation $\zeta(\vk)$. The isotropic
power spectrum is defines the variance of the curvature perturbations
as
\begin{eqnarray}\label{PowerSpectrum}
  \vev{\zeta(\vk)\zeta(\vk')}&=& (2\pi)^{3}\delta^{(3)}(\vk+\vk')
  P_\zeta(k) \nonumber\\
&\equiv& (2\pi)^{3}\delta^{(3)}(\vk+\vk')|\zeta(k)|^{2}_{k \ll aH}\,,
\end{eqnarray}
and is evaluated at a time when the amplitude of the mode has
converged on superhorizon scales ($k\ll aH$).

Expressed in terms of $N$ the Mukhanov-Sasaki equation becomes
\begin{equation}\label{Mukh}
  \frac{\mathrm{d}^{2}\zeta(k)}{\mathrm{d}N^{2}} + (3 + \epsilon - 2\eta)\frac{\mathrm{d}\zeta(k)}{\mathrm{d}N} + \frac{k^{2}}{a^{2}H^{2}}\zeta_{k} = 0\,,
\end{equation} 
from which it can also be seen that the amplitude of $\zeta(k)$ is
conserved on superhorizon scales.

The initial condition for integration of (\ref{Mukh}) is set when
$k\gg aH$ for each mode being solved for in which case the adiabatic
Bunch-Davies conditions can be assumed and the mode asymptotes to the
form
\begin{equation}\label{Initial_zeta}
  \zeta(k) \to \frac{e^{-ik\tau}}{2a\sqrt{k\epsilon}}\,,
\end{equation}
with $\tau$ the conformal time defined by $\mathrm{d}N/\mathrm{d}\tau
= aH$. The phase of $\zeta(k)$ is irrelevant and only the rate of change
for the initial condition on $\mathrm{d}\zeta_{k}/\mathrm{d}N$ is
required such that the value of $\tau$ at when the mode is normalised
need never be evaluated explicitly.

In the following (\ref{Mukh}) is integrated for a range of modes of
interest for observational comparison; $10^{-5} < k < 10^{-1}$ in
units of Mpc$^{-1}$. This is done for each flow trajectory drawn at
random in order to compare the resulting power spectrum to
observations via calculation of CMB angular power spectrum
\begin{equation}\label{cl}
  C_L = \int k^2\,dk \,P_\zeta(k) \,|\Delta_L(k)|^2\,,
\end{equation}
where $L$ here is the angular multipole and $\Delta_L(k,\eta_0)$ is
the multipole expanded radiation transfer function for the mode $k$
integrated to the present. The CMB angular power spectrum is evaluated
using modified version of {\tt CAMB} \cite{Lewis:1999bs} where
$P_\zeta(k|\flow{\ell})$ is used as input to (\ref{cl}) instead of the
conventional assumption
\begin{equation}
  k^3\,P_\zeta(k) = A_s \left(\frac{k}{k_\star}\right)^{n_s(k_\star)+\frac{1}{2} \frac{d
    n_s}{d\ln k}+\,...}\,,
\end{equation}
i.e. a power law with amplitude $A_s$ and spectral index given by
$n_s$ and higher derivative contributions . The power spectrum of
tensor modes $P_h(k|\flow{\ell})$ is calculated in a similar fashion
for the same range of wavenumbers and the tensor contribution to the
CMB angular power spectrum is also calculated. In this case the full
functional form of $P_h(k|\flow{\ell})$ replaces the parametrisation
in terms of the tensor-to-scalar ratio $r$ and tensor spectral index
$n_t$.

It is important to note that the numerical integration of mode
evolution provides {\sl exact} solutions (within numerical tolerances)
without use of any ``slow-roll'' assumptions. The results obtained are
therefore valid also in the case when the flow parameters are not
small as long as other necessary conditions of weak coupling and
linearity are satisfied\footnote{For further details of our numerical
  integration scheme see \cite{2013arXiv1311.3224H}}.

The remaining stochastic parameter is the initial condition for
$H$. This value only affects the overall amplitude of the perturbation
spectra and does not modify the solution for the flow
parameters. There is therefore more freedom in choosing where to
impose a normalisation. For this work a value for $\ln(\tilde H_{\rm
  inf})$ is drawn from a uniform distribution and used to normalise the
Hubble rate of the trajectory at a time when a chosen pivot scale $k_\star$ has
been outside the horizon for a few $e$-foldings i.e. when it's
amplitude has converged as
\begin{equation}
  H|_{k_\star\sim aH} = \frac{10^6}{4\pi\sqrt{2\pi}}\, \tilde H_{\rm inf}\,.
\end{equation} 
The value of $\tilde H_{\rm inf}$ is then related linearly to the final
amplitude of the curvature power spectrum.
\begin{equation}\label{H_normalisation}
  H(N_{c}) = 4\pi \sqrt{2\pi\epsilon(N_{c})}M_{pl}A_{s}
\end{equation}

\subsection{Non-Gaussianity}

The bispectrum is defined as
\begin{equation}\label{bispectrum}
  \!\!\!\!\!\vev{\zeta(\vk)\zeta(\vk')\zeta(\vk'')}= (2\pi)^{3}\delta^{(3)}(\vk+\vk'+\vk'')B(k,k',k'')\,,
\end{equation}
where momentum conservation forces $\vk$, $\vk'$, $\vk''$ to form a
closed triangle and isotropy implies $B(k,k',k'')$ only depends on
their magnitudes. It is convenient to work with a dimensionless
bispectrum, which is independent of the power spectrum amplitude,
often denoted as
\begin{eqnarray}
  \!\!\!\!\!\!\!f_{\mathrm{NL}}(k,k',k'') \equiv \frac{5}{6}B(k,k',k'')/\nonumber\\
   \left[P_\zeta(k)P_\zeta(k')+P_\zeta(k)P_\zeta(k'')+P_\zeta(k')P_\zeta(k'')\right]\,.
\end{eqnarray}
There are many ``type'' of $f_{\rm NL}$ with different weightings to
$P_\zeta(k)$ in the denominator while the above definition is
frequently called $f_{\rm NL}^{local}$.  The calculation of the
bispectrum relies on the ``in-in'' formalism to calculate
correlation-functions in time-dependent backgrounds for interacting
quantum fields.
\begin{equation}\label{in-in}
  \langle\zeta^{3}(t)\rangle = -i\int_{-\infty}^{t}\mathrm{d}t'\langle\left[\zeta^{3}(t), H_{\text{int}}(t')\right]\rangle\,,
\end{equation}
Just as in flat space when we express our fields as a sum of plane
waves (solutions to Klein-Gordon equation, Dirac equation etc.), here
we express $\zeta$ as a sum of solutions of (\ref{Mukh}).
\begin{equation}\label{zeta_quantised}
  \zeta(t,\mathbf{x}) = \int \frac{\mathrm{d}^{3}\mathbf{p}}{(2\pi)^{3}}\left(\zeta^{\,}_{\mathbf{p}}(t)\,a^{\,}_{\mathbf{p}} + \zeta^{*}_{-\mathbf{p}}(t)\,a^{\dagger}_{-\mathbf{p}}\right)\,e^{i\mathbf{p}\cdot\mathbf{x}}\,.
\end{equation}
$\zeta^{\,}_{\mathbf{p}}(t)$ by definition satisfies equation
(\ref{Mukh}) with initial condition (\ref{Initial_zeta}). The
interaction Hamiltonian $H_{\text{int}}(t')$ is obtained from
expanding the action for $\zeta$ to third order which produces cubic
interactions with time-dependent coupling
constants \cite{Maldacena:2002vr,Seery:2005gb,Hazra:2012yn,Funakoshi:2012ms,2013arXiv1311.3224H}.
\begin{eqnarray}\label{action_final}
  S_{3} &=& \!\!\int d^4x\, a^{3}\epsilon \left[\left(2\eta - \epsilon\right)\zeta \dot{\zeta}^{2} + \frac{1}{a^{2}}\epsilon\zeta(\partial\zeta)^{2}\right.\nonumber\\
  & &\!\!\!\!\left.  - (\epsilon - \eta)\zeta^{2}\partial^{2}\zeta - 2\epsilon\left(1 - \frac{\epsilon}{4}\right)\dot{\zeta}\partial_{i}\zeta\partial_{i}\partial^{-2}\dot{\zeta}\right.\nonumber\\
  & & \!\!\!\!\left. + \frac{\epsilon^{2}}{4}\partial^{2}\zeta\partial_{i}\partial^{-2}\dot{\zeta}\partial_{i}\partial^{-2}\dot{\zeta}\right]\,,
\end{eqnarray}

Using this expression for $H_{\text{int}}(t')$ in 
(\ref{in-in}) produces the following expression for $\fnl$
\begin{eqnarray}\label{fnl_prelim}
  \!\!\!\!\!\!\!f_{\mathrm{NL}} &=& \frac{1}{3|\zeta|^{4}}\times\nonumber\\
  &&\!\!\!\!\!\!\!\!\!\!\!\!{\cal I}\left[\zeta^{*3}\int_{N_{0}}^{N_{2}} dN\, f_{1}\zeta^{3} + f_{2}\zeta\zeta^{\prime 2}\right]\,,
\end{eqnarray}
where $\zeta = \zeta_{k}, \zeta_{\beta} = \zeta_{\beta k}$ and
$\zeta^{\prime} = \mathrm{d}\zeta/\mathrm{d}N$. The functions $f_{i}$
are given by
\begin{eqnarray}
  \!\!\!\!\!f_{1} & = & \frac{5k^{2}a\epsilon}{H}(2\eta - 3\epsilon)\,,\nonumber\\
  \!\!\!\!\!f_{2} & = & -5Ha^{3}\epsilon\left(4\eta - \frac{3}{4}\epsilon^{2}\right)\,,
\end{eqnarray}
$N_{0}$ and $N_{2}$ are $e$-folds when $\zeta_{k}$ is deep inside and
far outside the horizon respectively. The subtleties involved for
dealing with this integral numerically are fully explored in
\cite{2013arXiv1311.3224H}.

\section{Constraints on Hubble Flow trajectories}\label{sec:res}

\begin{figure}[t]
  \centering
  \includegraphics[width=3in]{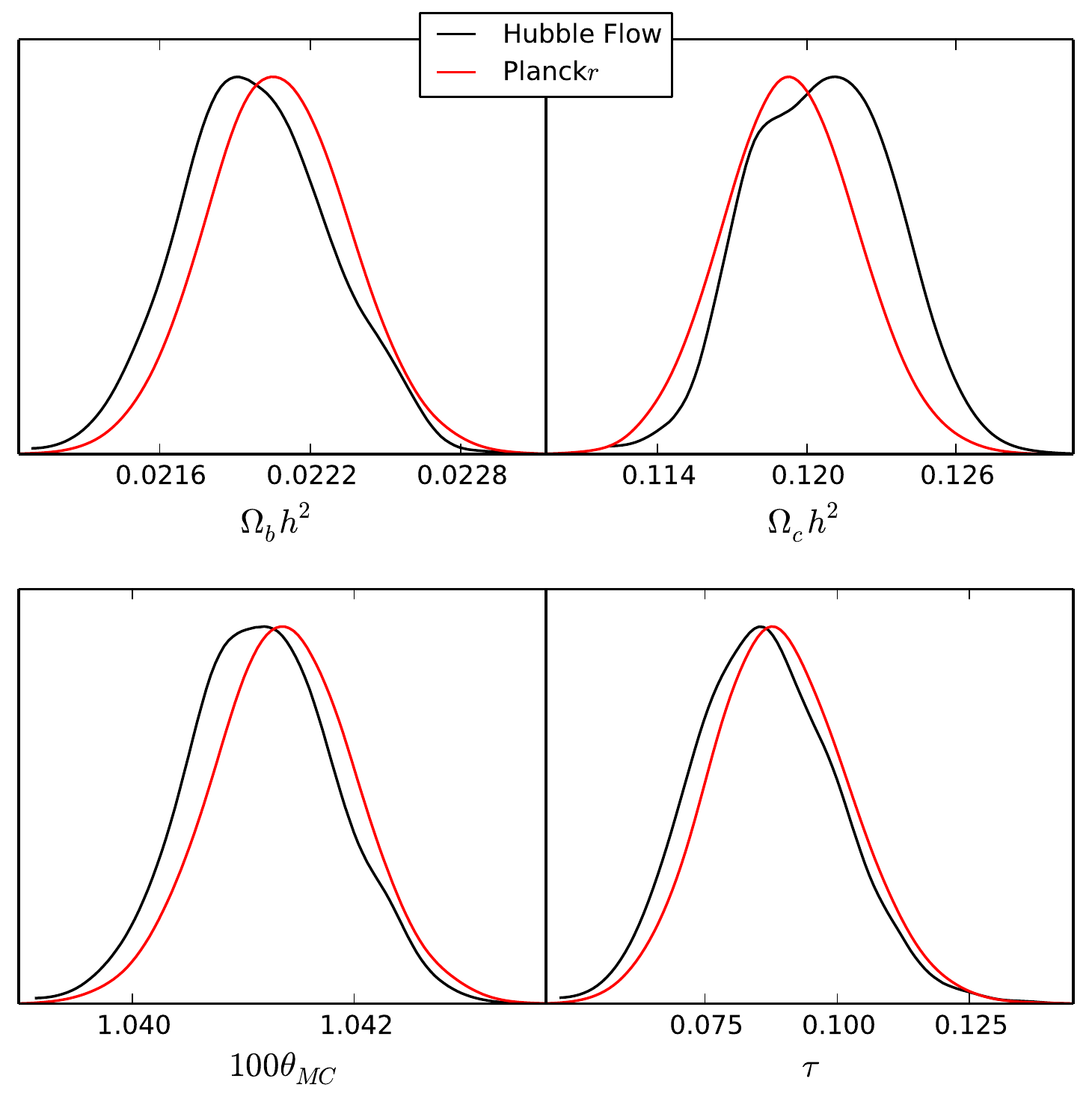}
  \caption{Comparison of 1d marginalised posteriors in the
    overlapping, parameters between the reference PLANCK$r$ run and
    the Hubble flow case with $\ell_{\rm max}=2$. there is no
    significant changes in the constraints as expected. The \planck\
    nuisance parameters are not shown but also shown no significant
    change in constraints.}
  \label{fig:acoustic}%
\end{figure}

\begin{figure}[t]
  \centering
  \includegraphics[width=3in]{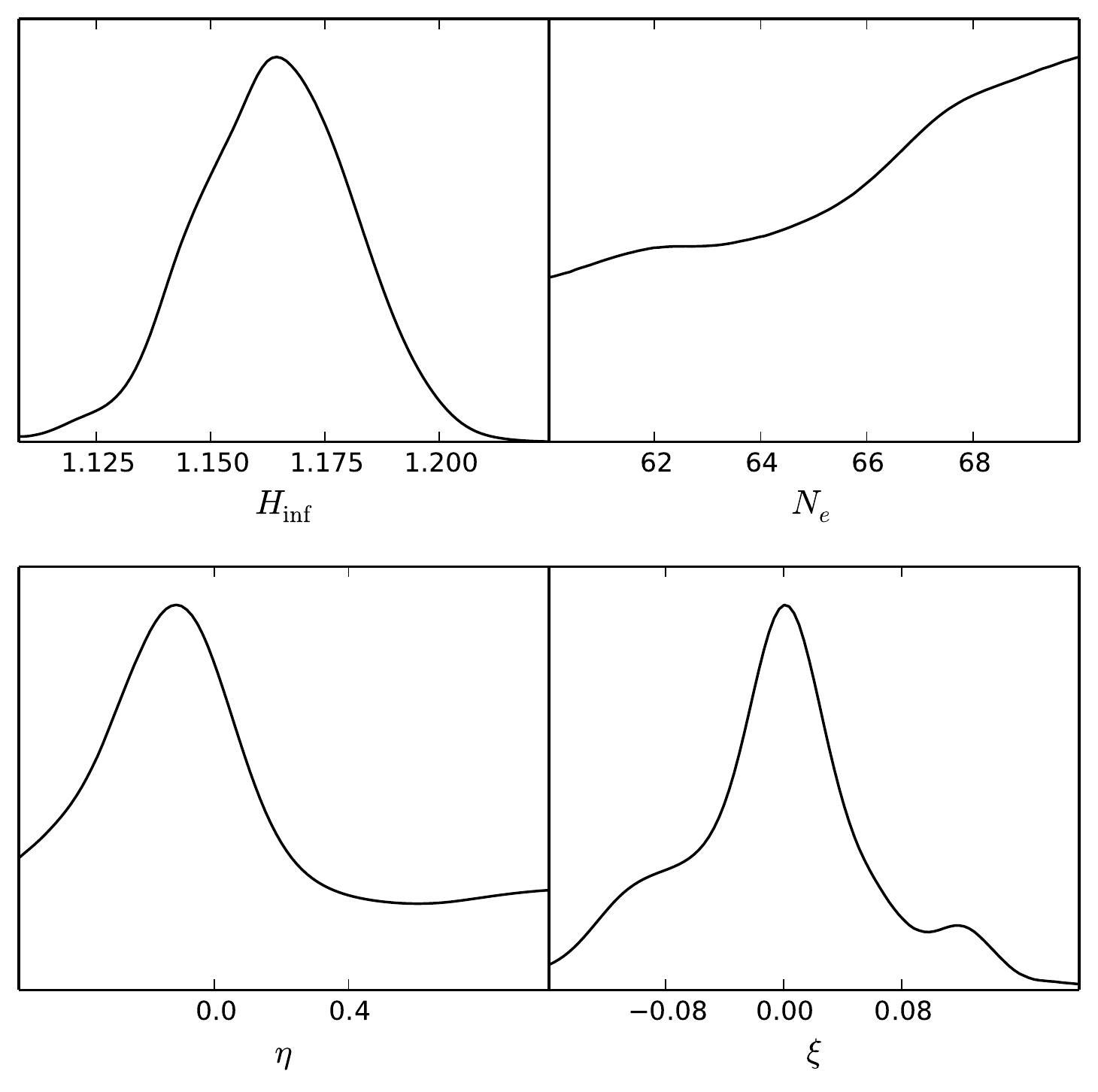}
  \caption{1d marginalised posteriors for the Hubble flow
    parameters. These parameters replace the conventional $A_s$,
    $n_s$, $r$, $n_t$, $dn_s/d\ln k$, etc. $H_{\rm inf}$ is equivalent
    to the scalar amplitude parameter $A_s$ and is well constrained,
    as expected, whereas the $e$-folds parameter $N_e$ is
    unconstrained. This is also expected since there is little
    sensitivity in the observable to the {\sl total} duration of
    inflation and $N_e$ can be regarded as an additional nuisance
    parameter. The flow parameters have posteriors
    peaked around $0$.}
  \label{fig:flow}%
\end{figure}

\subsection{Base parameters}

\begin{figure}[t]
  \centering
  \includegraphics[width=3.5in]{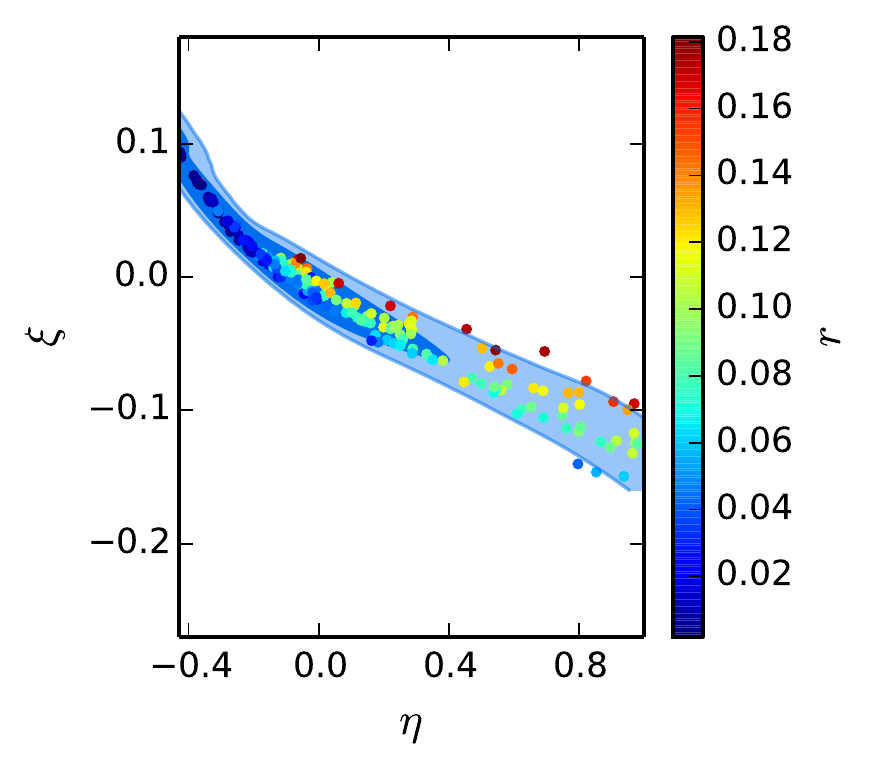}
  \caption{The 2d marginalised posterior for $\xi$ and $\eta$, the
    base flow parameters for the Hubble flow $\ell_{\rm max}=2$
    run. The contours are denote the 68\% and 95\% significance
    levels. The coloured scatter plot indicates the value of $r\sim 16\,
    \epsilon(k_\star)$ for each sample in the chain. The two base
    parameters are highly correlated and the unconstrained, large
    positive $\eta$ tail is correlated with larger values of $r$.}
  \label{fig:etaxi}%
\end{figure}

\begin{figure}[t]
  \centering
  \includegraphics[width=3.5in]{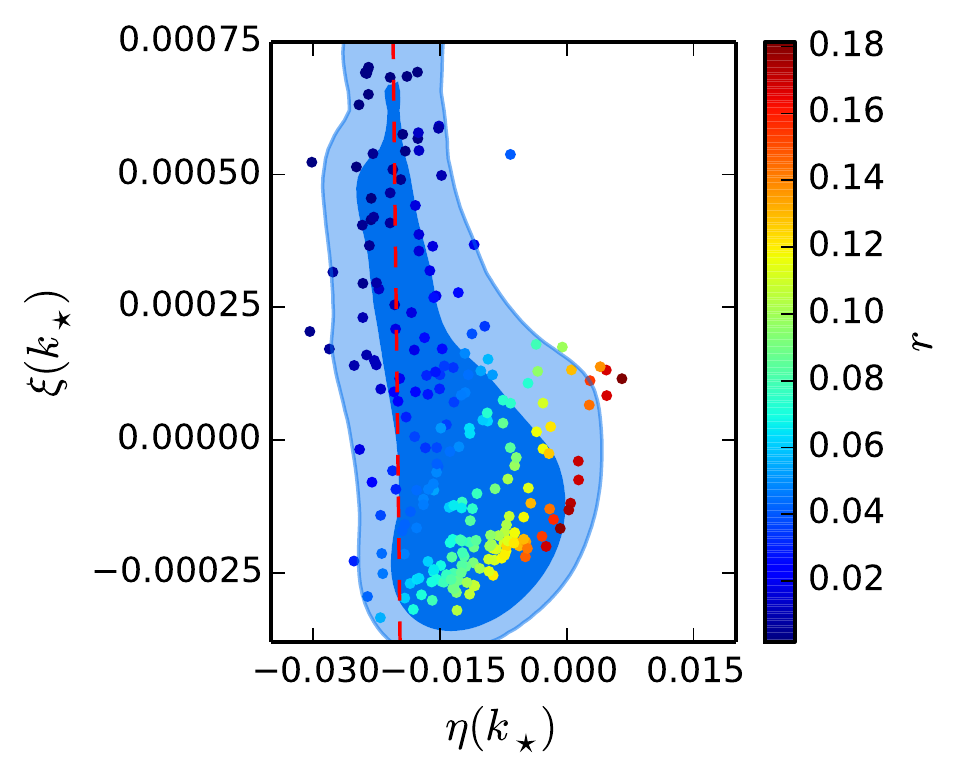}
  \caption{Same as Figure~\ref{fig:etaxi} but for $\xi$ and $\eta$
    values at the observationally relevant scale $k_\star$. The red
    (dashed) line indicates the expected value of $xi$ as $r\to 0$ as
    given by the second order slow-roll approximation.}
  \label{fig:etaxi_early}%
\end{figure}

Having defined a measure for generating random Hubble flow
trajectories one can now ask whether the resulting observables
i.e. scalar and tensor power spectra are compatible with observations
and/or gain constraints on the allowed space of flow parameters. To do
this the set of parameters defining the random trajectory $N_0$,
$H_{\rm inf}$, and $\flow{1}$, $\flow{2}$, ..., $\flow{\ell_{\rm
    max}}$ can be used as base parameters in an MCMC exploration of
the likelihood of CMB observations. In this case the set of flow
parameters replaces the conventional parametrisation of scalar and
tensor primordial power i.e. $A_s$, $n_s$, $dn_s/d\ln k$, etc., and
$r$, $n_t$, $dn_t/d\ln k$, etc.

Here, the {\tt CosmoMC} \cite{Lewis:2002ah} code is used, together with a
modified version of {\tt CAMB}, to explore the likelihood of the
Hubble flow parameters with respect to CMB observations. The parameter
set used in the exploration in this case is the combination of
radiation transfer parameters $\omega_b$, the physical density of
baryons, $\omega_c$, the physical density of cold dark matter,
$\theta_MC$, the angular diameter distance parameter used by ${\tt
  CosmoMC}$ \cite{2013arXiv1303.5076P}, and $\tau$, the optical depth
parameter, and the set of flow parameters $N_0$, $H_{\rm inf}$, and
$\flow{0}$, $\flow{1}$, ..., $\flow{\ell_{\rm max}}$ . The flow
parameters only affect the primordial scalar and tensor spectra and
are therefore probed as {\sl fast} parameters in {\tt CosmoMC}
runs. In practice at each step in the MCMC we compute the trajectory
resulting from the set of proposed flow parameters and then
numerically evaluate the corresponding scalar and tensor power
spectra. We also evaluate numerically the value of $\fnl$, the
dimensionless amplitude of the bispectrum in the equilateral
configuration, for the pivot scale.

For CMB observations, the latest \planck\ temperature only results
\cite{2013arXiv1303.5075P} are used together with \WMAP\ polarisation
measurements. In all the runs described in this work the \planck\
likelihood settings and nuisance parameters are set as in the standard
``PLANCK+WP'' combination ( see \cite{2013arXiv1303.5076P} for details). The
PLANCK+WP ``{\tt base r planck lowl lowLike}'' (abbreviated to
PLANCK$r$ in the following) MCMC chains \cite{Lewis:2002ah} using the
conventional parametrisation $A_s$, $n_s$ for the primordial scalar
spectrum with a tensor extension parametrised solely by $r$ can be
used as a reference run to compare with the results reported
below\footnote{For the tensor spectral index the inflationary
  consistency relation is used to treat it is as a function of
  $n_s$.}.

The conventional parameters $A_s$, $n_s$, $r$, etc., can be calculated
directly from the power spectra obtained by numerical integration of
the mode equations and can then be treated as {\sl derived} parameters
for each accepted flow trajectory in the MCMC chains. $\fnl$ can also
be treated as a {\sl derived} parameter to gain insight into the level
of non-Gaussianity preferred by the current data in the context of
random Hubble flow proposal.  It is instructive to visualise how the
Hubble flow proposal density used here projects into the space of
derived parameters. Figure~\ref{fig:proposal} shows the scatter of
trajectories in the $n_s$-$r$ and $n_s$-$\fnl$ planes. The derived
quantities are evaluated from the numerically obtained spectra at a
pivot scale $k_\star=0.05$ Mpc$^{-1}$. The points are also colour
coded according to the random value of $\eta$, $\xi$, and $N_0$ used
to generate the trajectory. The value of the random flow parameters is
highly correlated with the resulting values of $n_s$, $r$, and $\fnl$
and the scatter shows a strong ``inflationary'' attractor
\cite{Kinney:1997ne,Easther:2002rw,Liddle:2003py,2005PhRvD..72h3520C}. The attractor overlaps the PLANCK$r$ constraints for
the $n_s$-$r$ combination in a corner of the region between the 68\%
and 95\% contours.

\begin{table}[t]
  \caption{Parameter constraints from the marginalised posteriors for
    both Hubble flow $\ell_{\rm max}=2$ and PLANCK$r$ runs. Parameters
    marked with $^\dagger$ are derived ones in the Hubble flow run. Upper
    limits are 95\% significance values.}          
  \label{table:params}     
  \centering                         
  \begin{tabular}{lrr }       
    \hline\hline                
    $$ & Hubble Flow &PLANCK$r$\\
    \hline
    \\[0.1cm]
    $\Omega_b h^2$&$0.02198^{+0.00028}_{0.00032}$  & $0.02207^{+0.00028}_{-0.00028}$   \\ 
    $\Omega_b h^2$&$0.1206^{+0.0028}_{-0.0031}$ &$0.1193^{+0.0026}_{-0.0026}$\\
    $100\theta_{MC}$&$1.04117^{+0.00063}_{-0.00069}$ &$1.04137^{+0.00063}_{-0.00063}$\\
    $\tau$&$0.087^{+0.013}_{-0.015}$ &$0.089^{+0.012}_{-0.014}$\\
    $H_{\rm inf}$&$1.164^{+0.017}_{-0.017}$&-\\
    $\log(10^{10}A_s)$&-&$3.09^{+0.024}_{-0.027}$\\
    $^\dagger n_s$&$\,0.9579^{+0.0072}_{-0.0090}$&$0.9623^{+0.0075}_{-0.0075}$\\
    $^\dagger r$&$< 0.143$&$<0.126$\\
    $^\dagger \fnl$&$-0.0205^{+0.0037}_{-0.0057}$&-\\
    \\[0.1cm]
    \hline                                  
  \end{tabular}
\end{table}

We consider a Hubble flow system with $\ell_{\rm max}=2$ (i.e. including
$\epsilon$, $\eta$, and $\xi$) for the MCMC exploration. This allows
potentials that include up to order $6$ polynomials in $\phi$. The
uniform priors chosen for this run are shown in
table~\ref{table:params} together with a description of each base and
derived parameter. The run uses seven base parameters which is the
same number used for the conventional PLANCK$r$ run. The \planck\
nuisance parameters are omitted for brevity.

The chains are run until the $R^{-1}$ convergence parameter
\cite{Lewis:2002ah} falls below 0.1. Figure~\ref{fig:acoustic} shows the
resulting 1-dimensional marginalised posterior distribution for the
conventional parameters that determine the form of the radiation
perturbation transfer functions. These shared by both PLANCK$r$ 
and the Hubble flow runs. The marginalised
posteriors are very similar between the two runs indicating that there
is no {\sl tension} in the transfer parameters with respect to how the
primordial perturbation spectrum is sampled. 

Figure~\ref{fig:flow} shows the marginalised posteriors for the flow
parameters that do not have counterparts in the conventional runs. The
overall amplitude is tightly constrained as expected - it takes the
same role as the conventional amplitude $A_s$. The total number of
$e$-folds is unconstrained and acts an an extra nuisance parameter
which is marginalised in the given interval. The two flow parameters
whose values are allowed to vary at the end of inflation, $\eta$ and
$\xi$ have posteriors that are peaked around zero. The $\xi$ parameter
is also well constrained with respect to its uniform prior. Positive
values of $\eta$ are unconstrained and the posterior approaches a
uniform distribution that extends to the $\eta=1$ limit of the uniform
prior. 

The two Hubble flow parameters are highly correlated as seen in
Figure~\ref{fig:etaxi}. The large $\eta$, negative $\xi$ tail however
is correlated with larger values of $r\sim 16\,\epsilon(k_\star)$ and
therefore lower upper limits on the tensor-to-scalar ratio will help
in eliminating the large $\eta$ tail and break the degeneracy. The
resulting 2d posterior for the hubble flow parameters at observable
scales can be seen in Figure~\ref{fig:etaxi_early} that also
includes a line indicating the consistency of $\eta$ and $\xi$ with
the slow-roll limit expression in the limit that  $r\sim 16\,
\epsilon(k_\star) \to 0$ and $n_s-1\to -0.04$.

Trajectories with positive $\eta$ values, and hence positive
derivative in $\epsilon$ at the end of inflation approach the
slow--roll limit very quickly as they are evolved backwards towards
the observable window. These therefore almost always result in
acceptable values for e.g. $n_s$, $r$, etc. Negative values of $\eta$
are cutoff by the data at $\eta\sim -0.42$. The reason for this strong
cutoff can be seen in Figure~\ref{fig:traj} where trajectories with
larger negative values of $\eta$ at the end of inflation approach the
slow-roll limit much slower, giving values of $|n_s-1|$ that are
typically larger and therefore in disagreement with observations.

The best-fit sample in the chain for the $\ell_{\rm max}=2$ Hubble
flow run has a negative log-likelihood $-\ln L\equiv {\cal L}=4903.2521$ compared
to the PLANCK$r$ one of ${\cal L}=4904.3370$ giving a better
fit by $\Delta {\cal L} = 1.085$. The two runs have a comparable number
of degrees of freedom since the $N_e$ can be considered as an
additional nuisance parameter. The marginalised constraints on
parameters in both Hubble flow and PLANCK$r$ run are compared in Table~\ref{table:params}.

\subsection{Derived parameters}

\begin{figure}[t]
  \centering
  \includegraphics[width=3in]{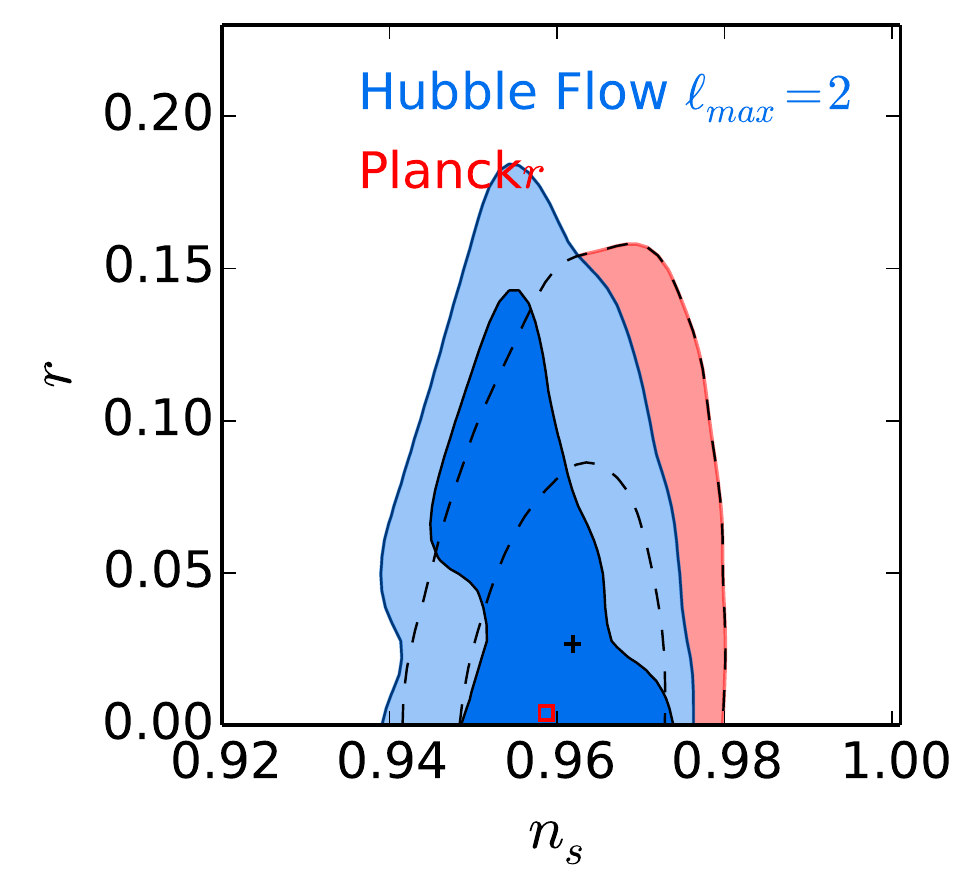}
  \caption{The 2d marginalised posterior for $n_s$ and $r$. These are
    derived at the pivot scale $k_\star$ in the Hubble flow case. The
    same contours for the PLANCK$r$ run are shown for comparison. The
    cross and square indicate the position of the best-fit sample for
    the Hubble flow and PLANCK$r$ run respectively. The Hubble flow
    case prefers higher values of $r$ due to the proposal density
    peaking at $r\sim0.075$ for acceptable values of $n_s$.}
  \label{fig:nsr}%
\end{figure}

It is useful to compare the marginalised posteriors in the derived $r$
and $n_s$ parameters between the Hubble flow run and the conventional
PLANCK$r$ case. Figure~\ref{fig:nsr} shows the 2d marginalised
constraints for this combination together with their respective
best-fit sample location.  The Hubble flow case prefers higher values
of $r$ due to the proposal density peaking at $r\sim0.075$ for
acceptable values of $n_s$. Constraints on $n_s$ are similar in both
cases although the Hubble flow constraints disfavour relatively large
values of $n_s$ compared to PLANCK$r$. 

\begin{figure}[t]
  \centering
  \includegraphics[width=3in]{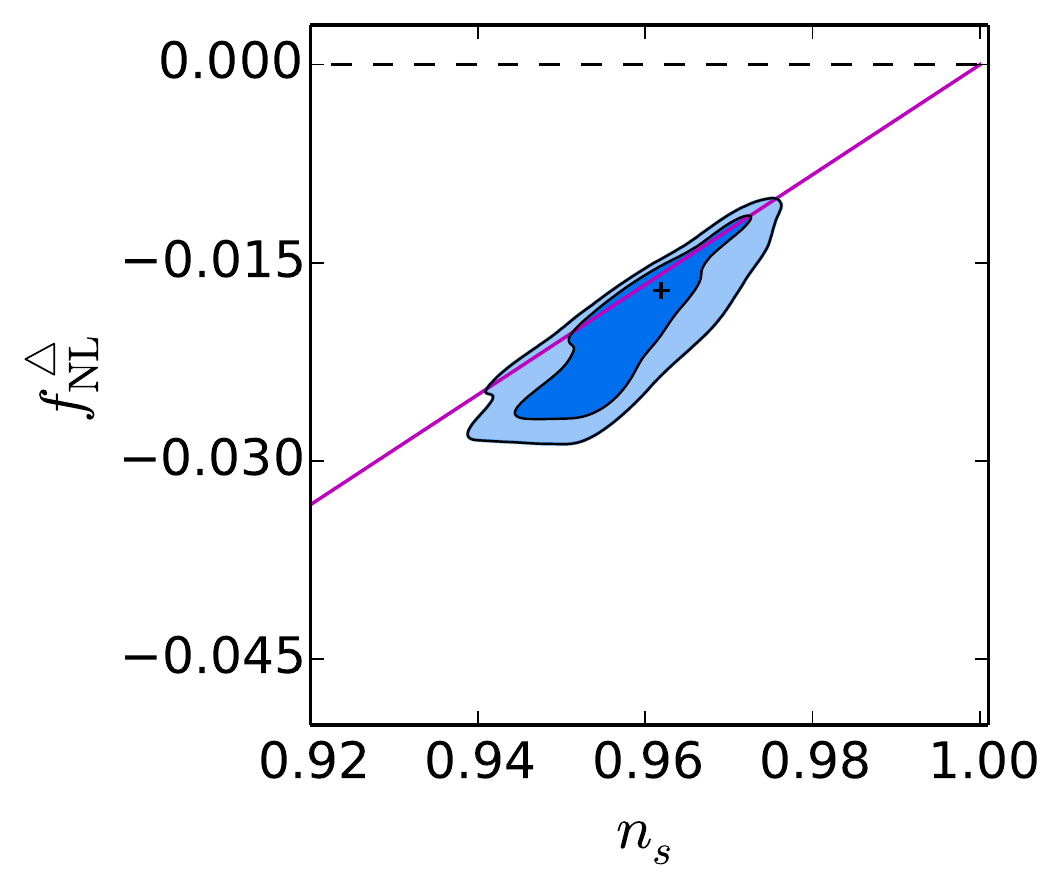}
  \caption{The 2d marginalised posterior for $n_s$ and $\fnl$ at the
    picot scale $k_\star$. The cross indicates the location of $\fnl$
    of the best-fit sample. The line shows the slow-roll consistency
    condition $\fnl\approx 5(n_s-1)/12$.}
  \label{fig:nsfnl}%
\end{figure}

A novel feature of this method is that existing data {\sl already}
constrains the possible values of $\fnl$. This is simply due to the
fact that each trajectory has a non-vanishing bispectrum and there
fore the data will constrain this degree of freedom
too. Figure~\ref{fig:nsfnl} shows the 2d marginalised constraints in
the $n_s$ {\sl vs} $\fnl$ plane. Since most of the trajectories are in
the slow--roll regime when the pivot scale $k_\star$ is leaving the
horizon the posterior for $\fnl$ agree well with the limiting
consistency condition $\fnl\approx 5(n_s-1)/12$
\cite{Maldacena:2002vr,2013arXiv1311.3224H}. This result, of course,  should not be interpreted as a
detection of non-Gaussianity but rather as an indication of what
amplitudes of the bispectrum are consistent with the general single
field inflationary solutions for a Hubble flow system with $\ell_{\rm
  max}$. If measurements of primordial non-Gaussianity ever reach the
sensitivity to constrain the level of $\fnl\sim 10^{-2}$ then the
measurement will provide a fundamental consistency check for single
field inflation.

\begin{figure}[t]
  \centering
    \includegraphics[width=3.5in]{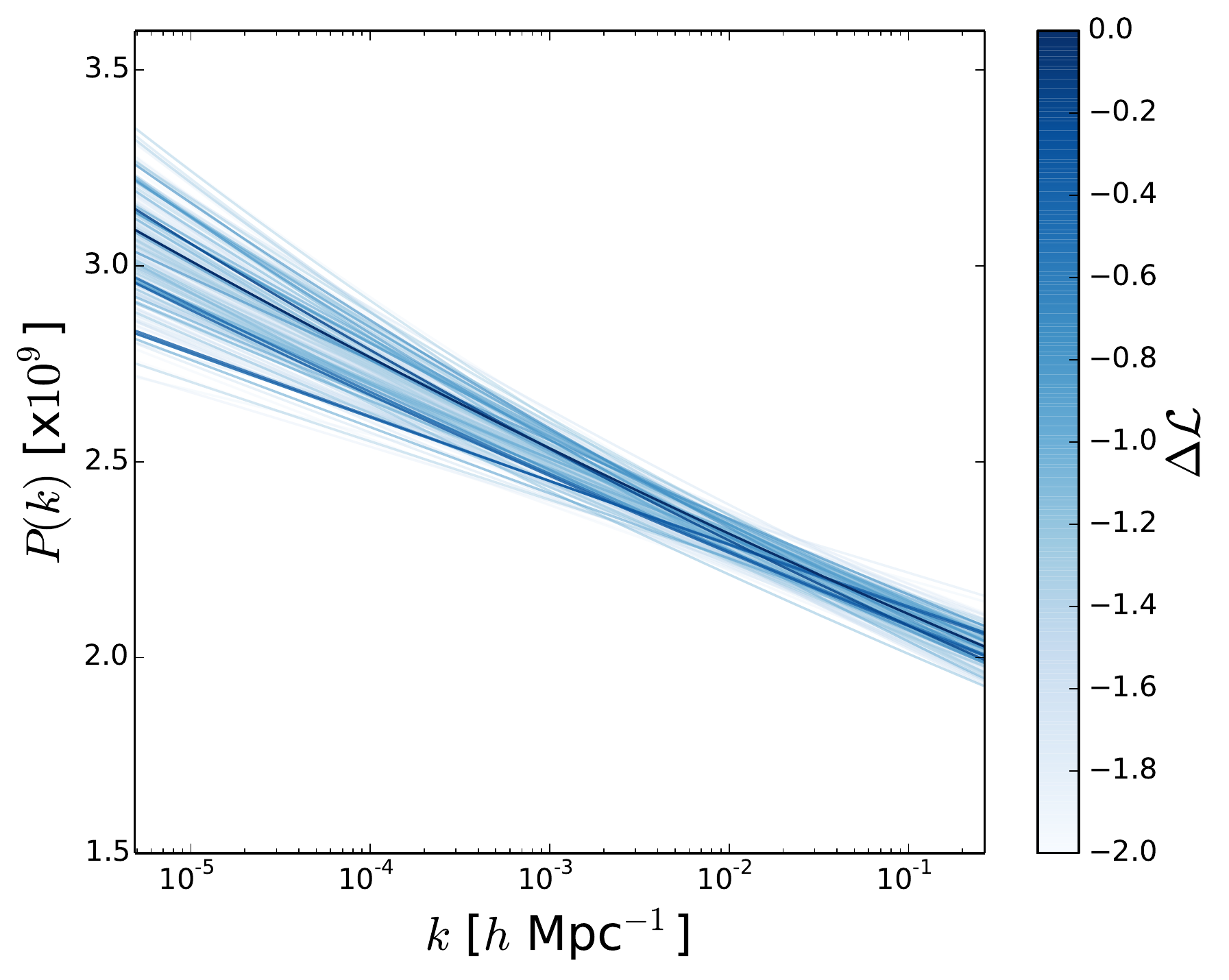}
    \caption{The primordial curvature power spectra for all samples s
      within $\Delta {\cal L}=2$ of the best-fit sample. This is an
      indication of all spectra allowed within the 95\% significance
      level. Each spectrum is colored and weighted on a scale given by
      $\Delta {\cal L}$ and $1/(1-\Delta {\cal L})$ respectively in
      order to emphasise the best-fitting spectra.  }
  \label{fig:ps}%
\end{figure}

In Figure~\ref{fig:ps} we also show the approximately 200 best-fit
power spectra in the chains. The spectra are colored and weighted by
their $\Delta {\cal L}$ with respect to the best-fit sample to
emphasise the best fitting curves. The best-fitting spectra are very
close to power laws with respect to $\ln k$. The best fitting spectra
have very similar normalisations at the pivot scale $k_\star=0.05 h$
Mpc$^{-1}$ as the normalisation of the primordial spectrum is once of
the best constrained parameters.

\subsection{Inflaton  potential}

\begin{figure*}[t]
  \centering
\begin{tabular}{ll}
    \includegraphics[width=3.5in]{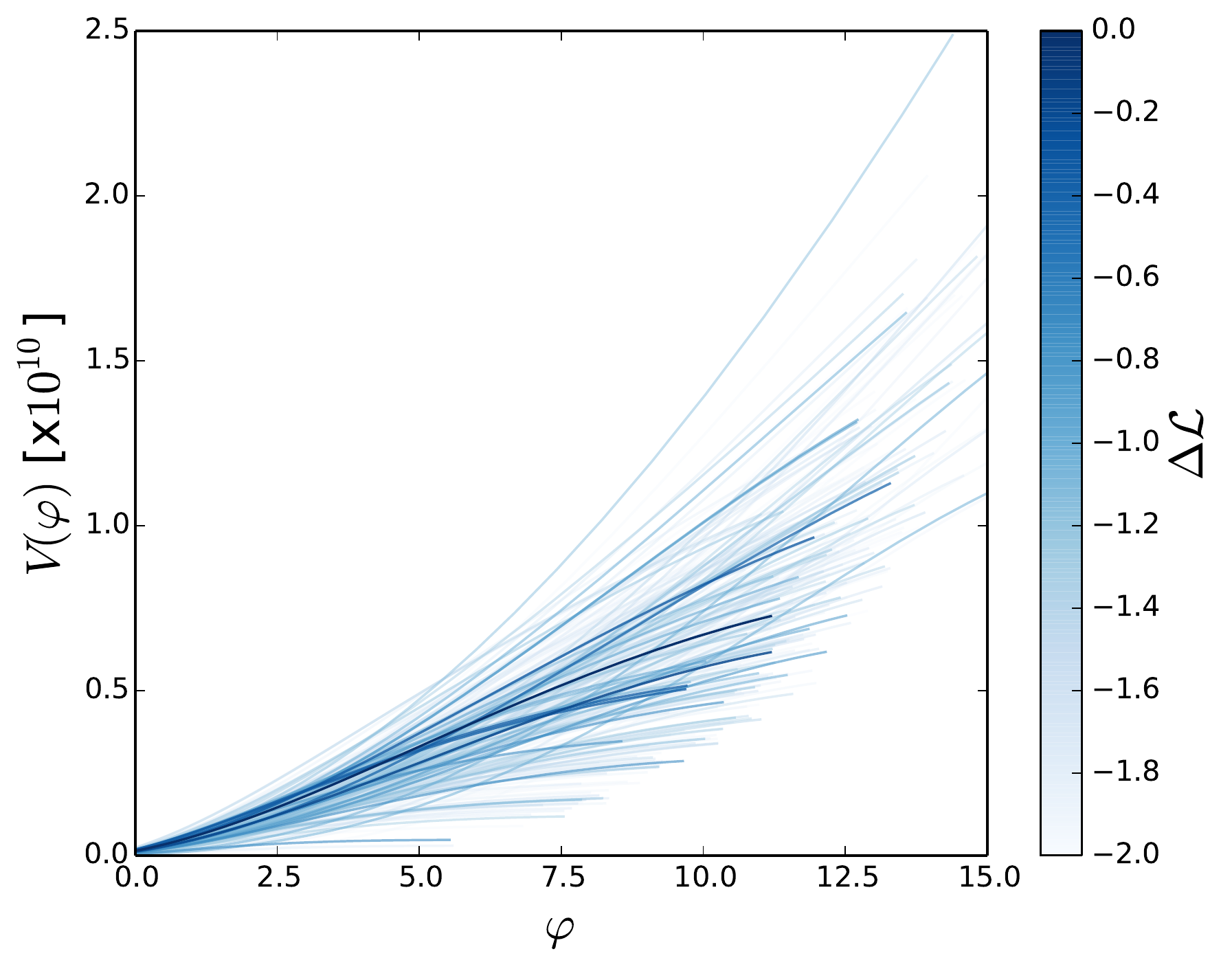}&
    \includegraphics[width=3.5in]{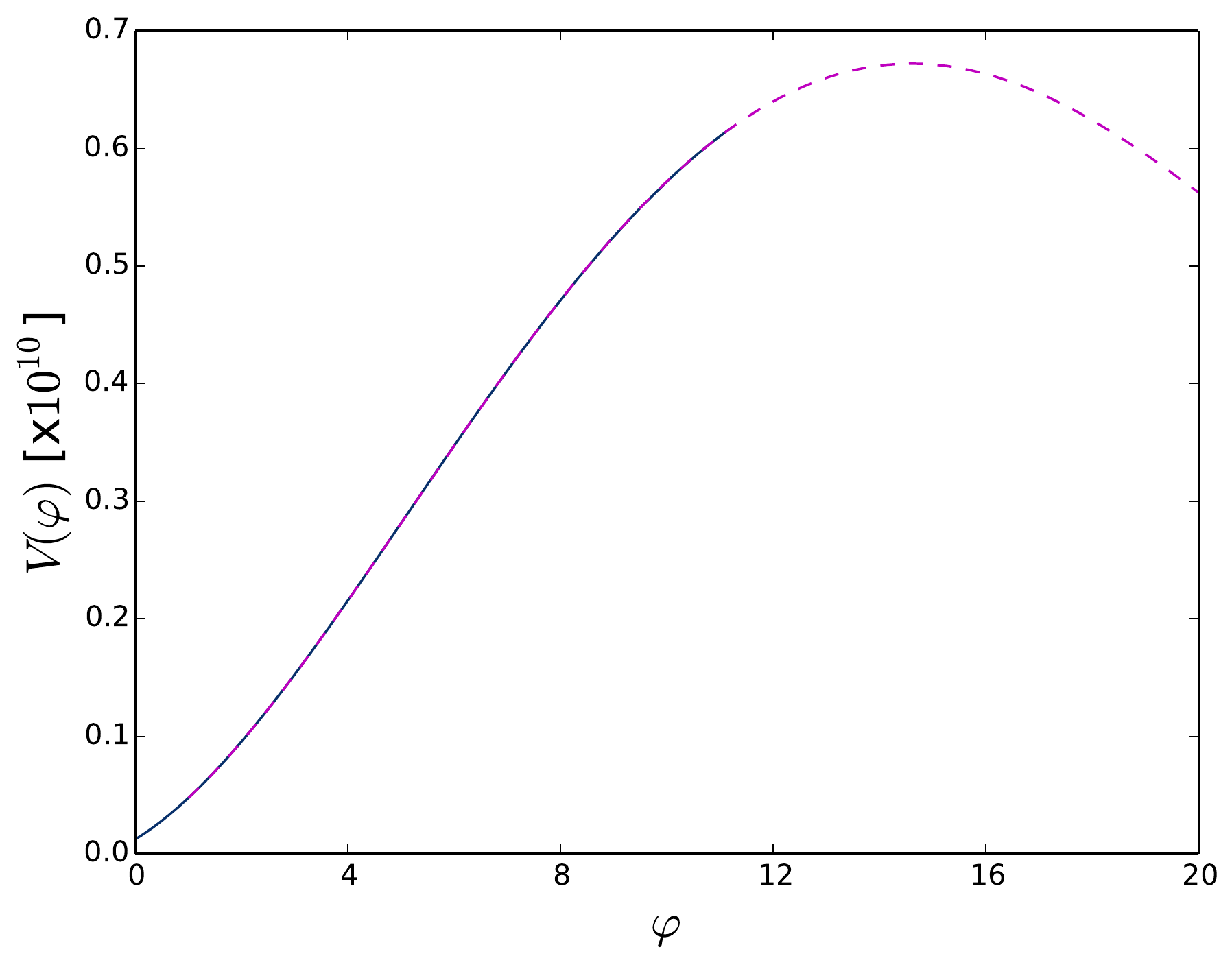}
  \end{tabular}
  \caption{{\sl Left}: All sampled potentials within $\Delta {\cal
      L}=2$ of the best-fit sample. This is an indication of all
    potential shapes and normalisations allowed with the 95\%
    significance level. The $x$-axis shows the change in $\phi$ from
    the final value where inflation ends ($\varphi\equiv\Delta
    \phi=0$). Both $\phi$ and $V$ are in units of $M_{\rm pl}=1$. The
    weighting of curves is the same as in Figure~\ref{fig:ps}. {\sl
      Right}: The best-fit sample potential (solid) and its 4$^{\rm
      th}$-order polynomial fit (dashed). }
  \label{fig:pots}%
\end{figure*}

Each trajectory in the MCMC chain yields an individual potential and
we can therefore translate directly the constraints on our base
parameters into the space of allowed potentials using
(\ref{potential}). For the $\ell_{\rm max}=2$ Hubble flow run the
best-fit potential is one given by $\eta\sim\xi\sim 0$ i.e. with small
curvature. Figure~\ref{fig:pots} shows all the potentials in the MCMC
chain that have $\Delta {\cal L}=2$ with respect to the best-fit
sample. There are some 200 samples within this range. Each potential
is weighted by its $\Delta {\cal L}$ value so the darkest curves are
the most likely.

The range in
$\eta$ probed by the sample is large and extends from $\eta\sim-0.4$
to $\eta\sim 1$. This translates to potentials that are both convex
and concave, and those that include an inflection point. This is
simply a feature of the degeneracy in the contribution from both
$\epsilon$ and $\eta$ to the scalar tilt $n_s$, the only shape
spectral parameter, aside from amplitude, that has been constrained so
far. If $r$ were to be detected in future it would help to constrain the sign of
the curvature of the potential in the observable regime
($\Delta\phi\sim 0$). 

An $n^{\rm th}$-order polynomial fit to the best-fit sample potential
converges for $n=5$ and gives a potential $V(\varphi)$ 
\begin{equation}
  V(\varphi) = V_0\left(1+\sum_{n=1}^{n=4}\lambda_n\,\varphi^n \right)\,,
\end{equation}
with $\varphi=\Delta \phi$, and $V_0=1.50\times 10^{-12}$, $\lambda_1 = 2.20$, $\lambda_2=0.66$,
$\lambda_3=-6.00\times 10^{-2}$, $\lambda_4=1.78\times
10^{-3}$, and $\lambda_5=-1.98\times
10^{-5}$. The best-fit sample potential and the polynomial fit are
shown in the left panel of Figure~\ref{fig:pots}.  The coefficients $\lambda_n$ for $n\le 4$
converge for higher order fits with $n>4$ and the potential does not
change appreciably in the interval $\varphi=0 \to {\cal O}(10)$. Note that given (\ref{potential}) an
$\ell_{\rm max}=2$ flow system allows for potentials that include
terms up to $\phi^6$.
\section{Discussion}\label{sec:disc}

We have obtained constraints on generalised, single field inflation
trajectories using the Hamilton-Jacobi formalism. The Hubble flow
system was used as base parameters in an MCMC exploration of the
likelihood with respect to the latest CMB data. This allowed us to
obtain marginalised posteriors on the flow parameters that define the
evolution of the Hubble parameter $H(N)$ as a function of $e$-folds
$N$ during inflation. Alternatively, the constraints can be viewed as
a selection in the space of inflaton potentials $V(\phi)$. 

Our method also includes the numerical calculation of primordial
bispectra and we obtained {\sl predictions} based on current data of
consistent bispectrum amplitude $f_{\rm NL}$ for the equilateral case.

Further exploration will be left for future work. In particular it
will be of interest to extend the system to higher $\ell_{\rm max}$ to
allow for more structure in the trajectories. This is currently
limited by the fact that the highly correlated space of HSR parameters
result in a very inefficient MCMC exploration. More work to explore
the likelihood more efficiently or defining new sets of HSR
parameters may help in extending this line of work to systems with
higher $\ell_{\rm max}$.

Future data from CMB and also large scale structure will also provide
deeper probes of non-Gaussianity which will provide tighter
constraints in the space of trajectories. This will be particularly
important if the discovery and characterisation of tensor modes will
turn out to elude future CMB polarisation measurements due to
foreground contamination. In that case non-Gaussianity measurements
will possibly provide the only way to break shape degeneracies and
reveal the precise form of the inflaton potential over the range of
scales accessible to observations.

\begin{acknowledgements}
  JSH is supported by a STFC studentship. CRC and JSH acknowledge the
  hospitality of the Perimeter Institute for Theoretical Physics  and
  the Canadian Institute for Theoretical Astrophysics where
  some of this work was carried out.
\end{acknowledgements}

\bibliography{ms}

\begin{thebibliography}{21}
\expandafter\ifx\csname natexlab\endcsname\relax\def\natexlab#1{#1}\fi
\expandafter\ifx\csname bibnamefont\endcsname\relax
  \def\bibnamefont#1{#1}\fi
\expandafter\ifx\csname bibfnamefont\endcsname\relax
  \def\bibfnamefont#1{#1}\fi
\expandafter\ifx\csname citenamefont\endcsname\relax
  \def\citenamefont#1{#1}\fi
\expandafter\ifx\csname url\endcsname\relax
  \def\url#1{\texttt{#1}}\fi
\expandafter\ifx\csname urlprefix\endcsname\relax\def\urlprefix{URL }\fi
\providecommand{\bibinfo}[2]{#2}
\providecommand{\eprint}[2][]{\url{#2}}

\bibitem[{\citenamefont{{Planck Collaboration}
  et~al.}(2013{\natexlab{a}})\citenamefont{{Planck Collaboration}, {Ade},
  {Aghanim}, {Armitage-Caplan}, {Arnaud}, {Ashdown}, {Atrio-Barandela},
  {Aumont}, {Baccigalupi}, {Banday} et~al.}}]{2013arXiv1303.5062P}
\bibinfo{author}{\bibnamefont{{Planck Collaboration}}},
  \bibinfo{author}{\bibfnamefont{P.~A.~R.} \bibnamefont{{Ade}}},
  \bibinfo{author}{\bibfnamefont{N.}~\bibnamefont{{Aghanim}}},
  \bibinfo{author}{\bibfnamefont{C.}~\bibnamefont{{Armitage-Caplan}}},
  \bibinfo{author}{\bibfnamefont{M.}~\bibnamefont{{Arnaud}}},
  \bibinfo{author}{\bibfnamefont{M.}~\bibnamefont{{Ashdown}}},
  \bibinfo{author}{\bibfnamefont{F.}~\bibnamefont{{Atrio-Barandela}}},
  \bibinfo{author}{\bibfnamefont{J.}~\bibnamefont{{Aumont}}},
  \bibinfo{author}{\bibfnamefont{C.}~\bibnamefont{{Baccigalupi}}},
  \bibinfo{author}{\bibfnamefont{A.~J.} \bibnamefont{{Banday}}},
  \bibnamefont{et~al.}, \bibinfo{journal}{ArXiv e-prints}
  (\bibinfo{year}{2013}{\natexlab{a}}), \eprint{1303.5062}.

\bibitem[{\citenamefont{{Planck Collaboration}
  et~al.}(2013{\natexlab{b}})\citenamefont{{Planck Collaboration}, {Ade},
  {Aghanim}, {Armitage-Caplan}, {Arnaud}, {Ashdown}, {Atrio-Barandela},
  {Aumont}, {Baccigalupi}, {Banday} et~al.}}]{2013arXiv1303.5082P}
\bibinfo{author}{\bibnamefont{{Planck Collaboration}}},
  \bibinfo{author}{\bibfnamefont{P.~A.~R.} \bibnamefont{{Ade}}},
  \bibinfo{author}{\bibfnamefont{N.}~\bibnamefont{{Aghanim}}},
  \bibinfo{author}{\bibfnamefont{C.}~\bibnamefont{{Armitage-Caplan}}},
  \bibinfo{author}{\bibfnamefont{M.}~\bibnamefont{{Arnaud}}},
  \bibinfo{author}{\bibfnamefont{M.}~\bibnamefont{{Ashdown}}},
  \bibinfo{author}{\bibfnamefont{F.}~\bibnamefont{{Atrio-Barandela}}},
  \bibinfo{author}{\bibfnamefont{J.}~\bibnamefont{{Aumont}}},
  \bibinfo{author}{\bibfnamefont{C.}~\bibnamefont{{Baccigalupi}}},
  \bibinfo{author}{\bibfnamefont{A.~J.} \bibnamefont{{Banday}}},
  \bibnamefont{et~al.}, \bibinfo{journal}{ArXiv e-prints}
  (\bibinfo{year}{2013}{\natexlab{b}}), \eprint{1303.5082}.

\bibitem[{\citenamefont{{Kinney}}(1997)}]{Kinney:1997ne}
\bibinfo{author}{\bibfnamefont{W.~H.} \bibnamefont{{Kinney}}},
  \bibinfo{journal}{\prd} \textbf{\bibinfo{volume}{56}}, \bibinfo{pages}{2002}
  (\bibinfo{year}{1997}), \eprint{hep-ph/9702427}.

\bibitem[{\citenamefont{Easther and Kinney}(2003)}]{Easther:2002rw}
\bibinfo{author}{\bibfnamefont{R.}~\bibnamefont{Easther}} \bibnamefont{and}
  \bibinfo{author}{\bibfnamefont{W.~H.} \bibnamefont{Kinney}},
  \bibinfo{journal}{Phys.Rev.} \textbf{\bibinfo{volume}{D67}},
  \bibinfo{pages}{043511} (\bibinfo{year}{2003}), \eprint{astro-ph/0210345}.

\bibitem[{\citenamefont{Liddle}(2003)}]{Liddle:2003py}
\bibinfo{author}{\bibfnamefont{A.~R.} \bibnamefont{Liddle}},
  \bibinfo{journal}{Phys.Rev.} \textbf{\bibinfo{volume}{D68}},
  \bibinfo{pages}{103504} (\bibinfo{year}{2003}), \eprint{astro-ph/0307286}.

\bibitem[{\citenamefont{{Chongchitnan} and
  {Efstathiou}}(2005)}]{2005PhRvD..72h3520C}
\bibinfo{author}{\bibfnamefont{S.}~\bibnamefont{{Chongchitnan}}}
  \bibnamefont{and}
  \bibinfo{author}{\bibfnamefont{G.}~\bibnamefont{{Efstathiou}}},
  \bibinfo{journal}{\prd} \textbf{\bibinfo{volume}{72}}, \bibinfo{eid}{083520}
  (\bibinfo{year}{2005}), \eprint{astro-ph/0508355}.

\bibitem[{\citenamefont{Salopek and Bond}(1990)}]{PhysRevD.42.3936}
\bibinfo{author}{\bibfnamefont{D.~S.} \bibnamefont{Salopek}} \bibnamefont{and}
  \bibinfo{author}{\bibfnamefont{J.~R.} \bibnamefont{Bond}},
  \bibinfo{journal}{Phys. Rev. D} \textbf{\bibinfo{volume}{42}},
  \bibinfo{pages}{3936} (\bibinfo{year}{1990}),
  \urlprefix\url{http://link.aps.org/doi/10.1103/PhysRevD.42.3936}.

\bibitem[{\citenamefont{{Bennett} et~al.}(2013)\citenamefont{{Bennett},
  {Larson}, {Weiland}, {Jarosik}, {Hinshaw}, {Odegard}, {Smith}, {Hill},
  {Gold}, {Halpern} et~al.}}]{2013ApJS..208...20B}
\bibinfo{author}{\bibfnamefont{C.~L.} \bibnamefont{{Bennett}}},
  \bibinfo{author}{\bibfnamefont{D.}~\bibnamefont{{Larson}}},
  \bibinfo{author}{\bibfnamefont{J.~L.} \bibnamefont{{Weiland}}},
  \bibinfo{author}{\bibfnamefont{N.}~\bibnamefont{{Jarosik}}},
  \bibinfo{author}{\bibfnamefont{G.}~\bibnamefont{{Hinshaw}}},
  \bibinfo{author}{\bibfnamefont{N.}~\bibnamefont{{Odegard}}},
  \bibinfo{author}{\bibfnamefont{K.~M.} \bibnamefont{{Smith}}},
  \bibinfo{author}{\bibfnamefont{R.~S.} \bibnamefont{{Hill}}},
  \bibinfo{author}{\bibfnamefont{B.}~\bibnamefont{{Gold}}},
  \bibinfo{author}{\bibfnamefont{M.}~\bibnamefont{{Halpern}}},
  \bibnamefont{et~al.}, \bibinfo{journal}{\apjs}
  \textbf{\bibinfo{volume}{208}}, \bibinfo{eid}{20} (\bibinfo{year}{2013}),
  \eprint{1212.5225}.

\bibitem[{\citenamefont{Bunch and Davies}(1978)}]{Bunch:1978yq}
\bibinfo{author}{\bibfnamefont{T.}~\bibnamefont{Bunch}} \bibnamefont{and}
  \bibinfo{author}{\bibfnamefont{P.}~\bibnamefont{Davies}},
  \bibinfo{journal}{Proc.Roy.Soc.Lond.} \textbf{\bibinfo{volume}{A360}},
  \bibinfo{pages}{117} (\bibinfo{year}{1978}).

\bibitem[{\citenamefont{{Planck Collaboration}
  et~al.}(2013{\natexlab{c}})\citenamefont{{Planck Collaboration}, {Ade},
  {Aghanim}, {Armitage-Caplan}, {Arnaud}, {Ashdown}, {Atrio-Barandela},
  {Aumont}, {Baccigalupi}, {Banday} et~al.}}]{2013arXiv1303.5076P}
\bibinfo{author}{\bibnamefont{{Planck Collaboration}}},
  \bibinfo{author}{\bibfnamefont{P.~A.~R.} \bibnamefont{{Ade}}},
  \bibinfo{author}{\bibfnamefont{N.}~\bibnamefont{{Aghanim}}},
  \bibinfo{author}{\bibfnamefont{C.}~\bibnamefont{{Armitage-Caplan}}},
  \bibinfo{author}{\bibfnamefont{M.}~\bibnamefont{{Arnaud}}},
  \bibinfo{author}{\bibfnamefont{M.}~\bibnamefont{{Ashdown}}},
  \bibinfo{author}{\bibfnamefont{F.}~\bibnamefont{{Atrio-Barandela}}},
  \bibinfo{author}{\bibfnamefont{J.}~\bibnamefont{{Aumont}}},
  \bibinfo{author}{\bibfnamefont{C.}~\bibnamefont{{Baccigalupi}}},
  \bibinfo{author}{\bibfnamefont{A.~J.} \bibnamefont{{Banday}}},
  \bibnamefont{et~al.}, \bibinfo{journal}{ArXiv e-prints}
  (\bibinfo{year}{2013}{\natexlab{c}}), \eprint{1303.5076}.

\bibitem[{\citenamefont{{Horner} and
  {Contaldi}}(2013{\natexlab{a}})}]{2013arXiv1303.2119H}
\bibinfo{author}{\bibfnamefont{J.~S.} \bibnamefont{{Horner}}} \bibnamefont{and}
  \bibinfo{author}{\bibfnamefont{C.~R.} \bibnamefont{{Contaldi}}},
  \bibinfo{journal}{ArXiv e-prints}  (\bibinfo{year}{2013}{\natexlab{a}}),
  \eprint{1303.2119}.

\bibitem[{\citenamefont{{Horner} and
  {Contaldi}}(2013{\natexlab{b}})}]{2013arXiv1311.3224H}
\bibinfo{author}{\bibfnamefont{J.~S.} \bibnamefont{{Horner}}} \bibnamefont{and}
  \bibinfo{author}{\bibfnamefont{C.~R.} \bibnamefont{{Contaldi}}},
  \bibinfo{journal}{ArXiv e-prints}  (\bibinfo{year}{2013}{\natexlab{b}}),
  \eprint{1311.3224}.

\bibitem[{\citenamefont{Mukhanov and Chibisov}(1982)}]{mukhanov1982vacuum}
\bibinfo{author}{\bibfnamefont{V.}~\bibnamefont{Mukhanov}} \bibnamefont{and}
  \bibinfo{author}{\bibfnamefont{G.}~\bibnamefont{Chibisov}},
  \bibinfo{journal}{Zh. Eksp. Teor. Fiz} \textbf{\bibinfo{volume}{83}},
  \bibinfo{pages}{487} (\bibinfo{year}{1982}).

\bibitem[{\citenamefont{Sasaki}(1986)}]{Sasaki:1986hm}
\bibinfo{author}{\bibfnamefont{M.}~\bibnamefont{Sasaki}},
  \bibinfo{journal}{Prog.Theor.Phys.} \textbf{\bibinfo{volume}{76}},
  \bibinfo{pages}{1036} (\bibinfo{year}{1986}).

\bibitem[{\citenamefont{Lewis et~al.}(2000)\citenamefont{Lewis, Challinor, and
  Lasenby}}]{Lewis:1999bs}
\bibinfo{author}{\bibfnamefont{A.}~\bibnamefont{Lewis}},
  \bibinfo{author}{\bibfnamefont{A.}~\bibnamefont{Challinor}},
  \bibnamefont{and} \bibinfo{author}{\bibfnamefont{A.}~\bibnamefont{Lasenby}},
  \bibinfo{journal}{Astrophys. J.} \textbf{\bibinfo{volume}{538}},
  \bibinfo{pages}{473} (\bibinfo{year}{2000}), \eprint{astro-ph/9911177}.

\bibitem[{\citenamefont{Maldacena}(2003)}]{Maldacena:2002vr}
\bibinfo{author}{\bibfnamefont{J.~M.} \bibnamefont{Maldacena}},
  \bibinfo{journal}{JHEP} \textbf{\bibinfo{volume}{0305}}, \bibinfo{pages}{013}
  (\bibinfo{year}{2003}), \eprint{astro-ph/0210603}.

\bibitem[{\citenamefont{Seery and Lidsey}(2005)}]{Seery:2005gb}
\bibinfo{author}{\bibfnamefont{D.}~\bibnamefont{Seery}} \bibnamefont{and}
  \bibinfo{author}{\bibfnamefont{J.~E.} \bibnamefont{Lidsey}},
  \bibinfo{journal}{JCAP} \textbf{\bibinfo{volume}{0509}}, \bibinfo{pages}{011}
  (\bibinfo{year}{2005}), \eprint{astro-ph/0506056}.

\bibitem[{\citenamefont{Hazra et~al.}(2012)\citenamefont{Hazra, Sriramkumar,
  and Martin}}]{Hazra:2012yn}
\bibinfo{author}{\bibfnamefont{D.~K.} \bibnamefont{Hazra}},
  \bibinfo{author}{\bibfnamefont{L.}~\bibnamefont{Sriramkumar}},
  \bibnamefont{and} \bibinfo{author}{\bibfnamefont{J.}~\bibnamefont{Martin}}
  (\bibinfo{year}{2012}), \eprint{1201.0926}.

\bibitem[{\citenamefont{Funakoshi and Renaux-Petel}(2013)}]{Funakoshi:2012ms}
\bibinfo{author}{\bibfnamefont{H.}~\bibnamefont{Funakoshi}} \bibnamefont{and}
  \bibinfo{author}{\bibfnamefont{S.}~\bibnamefont{Renaux-Petel}},
  \bibinfo{journal}{JCAP} \textbf{\bibinfo{volume}{1302}}, \bibinfo{pages}{002}
  (\bibinfo{year}{2013}), \eprint{1211.3086}.

\bibitem[{\citenamefont{Lewis and Bridle}(2002)}]{Lewis:2002ah}
\bibinfo{author}{\bibfnamefont{A.}~\bibnamefont{Lewis}} \bibnamefont{and}
  \bibinfo{author}{\bibfnamefont{S.}~\bibnamefont{Bridle}},
  \bibinfo{journal}{Phys. Rev.} \textbf{\bibinfo{volume}{D66}},
  \bibinfo{pages}{103511} (\bibinfo{year}{2002}), \eprint{astro-ph/0205436}.

\bibitem[{\citenamefont{{Planck collaboration}
  et~al.}(2013)\citenamefont{{Planck collaboration}, {Ade}, {Aghanim},
  {Armitage-Caplan}, {Arnaud}, {Ashdown}, {Atrio-Barandela}, {Aumont},
  {Baccigalupi}, {Banday} et~al.}}]{2013arXiv1303.5075P}
\bibinfo{author}{\bibnamefont{{Planck collaboration}}},
  \bibinfo{author}{\bibfnamefont{P.~A.~R.} \bibnamefont{{Ade}}},
  \bibinfo{author}{\bibfnamefont{N.}~\bibnamefont{{Aghanim}}},
  \bibinfo{author}{\bibfnamefont{C.}~\bibnamefont{{Armitage-Caplan}}},
  \bibinfo{author}{\bibfnamefont{M.}~\bibnamefont{{Arnaud}}},
  \bibinfo{author}{\bibfnamefont{M.}~\bibnamefont{{Ashdown}}},
  \bibinfo{author}{\bibfnamefont{F.}~\bibnamefont{{Atrio-Barandela}}},
  \bibinfo{author}{\bibfnamefont{J.}~\bibnamefont{{Aumont}}},
  \bibinfo{author}{\bibfnamefont{C.}~\bibnamefont{{Baccigalupi}}},
  \bibinfo{author}{\bibfnamefont{A.~J.} \bibnamefont{{Banday}}},
  \bibnamefont{et~al.}, \bibinfo{journal}{ArXiv e-prints}
  (\bibinfo{year}{2013}), \eprint{1303.5075}.

\end{thebibliography}



\end{document}